\documentclass[aps,pra,twocolumn,superscriptaddress]{revtex4-2}

\usepackage{graphicx,subfigure}
\usepackage{amsmath,amssymb,bm,amsthm}
\usepackage{mathtools}
\usepackage{multirow}
\usepackage{textcomp}
\usepackage{float}
\usepackage{xcolor}
\usepackage[normalem]{ulem}
\usepackage{dsfont}
\usepackage{url}
\usepackage{mathrsfs}
\usepackage{physics}
\usepackage{tikz}
\usetikzlibrary{shapes, arrows, positioning}

\newcommand{\inp}[2]{\left \langle #1 \middle | #2 \right \rangle}
\newcommand{\opa}{\hat{a}}

\usepackage{hyperref}
\hypersetup{
    colorlinks,
    citecolor=black,
    filecolor=black,
    linkcolor=black,
    urlcolor=black
}

\begin{document}

\title{Quantum random access memory with transmon-controlled phonon routing}

\author{Zhaoyou Wang}
\thanks{These two authors contributed equally}
\email{zhaoyou@uchicago.edu}
\affiliation{Pritzker School of Molecular Engineering, University of Chicago, Chicago IL 60637, USA}

\author{Hong Qiao}
\thanks{These two authors contributed equally}
\email{hongqiao@uchicago.edu}
\affiliation{Pritzker School of Molecular Engineering, University of Chicago, Chicago IL 60637, USA}

\author{Andrew N. Cleland}
\email{anc@uchicago.edu}
\affiliation{Pritzker School of Molecular Engineering, University of Chicago, Chicago IL 60637, USA}
\affiliation{Center for Molecular Engineering and Material Science Division, Argonne National Laboratory, Lemont IL 60439, USA}

\author{Liang Jiang}
\email{liangjiang@uchicago.edu}
\affiliation{Pritzker School of Molecular Engineering, University of Chicago, Chicago IL 60637, USA}

\date{\today}

\begin{abstract}
    Quantum random access memory (QRAM) promises simultaneous data queries at multiple memory locations, with data retrieved in coherent superpositions, essential for achieving quantum speedup in many quantum algorithms.
    We introduce a transmon-controlled phonon router and propose a QRAM implementation by connecting these routers in a tree-like architecture.
    The router controls the motion of itinerant surface acoustic wave phonons based on the state of the control transmon, implementing the core functionality of conditional routing for QRAM.
    Our QRAM design is compact, supports fast routing operations, and avoids frequency crowding.
    Additionally, we propose a hybrid dual-rail encoding method to detect dominant loss errors without additional hardware, a versatile approach applicable to other QRAM platforms. Our estimates indicate that the proposed QRAM platform can achieve high heralding rates using current device parameters, with heralding fidelity primarily limited by transmon dephasing.
\end{abstract}

\maketitle

\emph{Introduction}\textemdash
The ability to store and manipulate information is fundamental to computing devices. In classical computers, random access memory (RAM) provides flexible access to data stored in an array of memory cells.
Similarly, quantum random access memory (QRAM) enables simultaneous retrieval of data in a coherent superposition at different memory locations~\cite{giovannetti2008a}, key to implementations of the quantum oracles used in many quantum algorithms, including searching~\cite{grover1996}, solving linear systems~\cite{harrow2009}, quantum machine learning~\cite{biamonte2017,ciliberto2018,lloyd2013,kerenidis2017,kerenidis2020} and quantum chemistry~\cite{cao2019,bauer2020,babbush2018}.
More precisely, given an n-bit string $j$ as input address, classical RAM retrieves the data bit $D_j=0$ or 1 stored at the $j$th memory location. In contrast, QRAM can query multiple addresses, mapping the output bus register from $\ket{0}_b$ to $\ket{D_j}_b$ conditioned on the state $\ket{j}_a$ of the input address register as $\sum_{j=1}^N \alpha_j \ket{j}_a \ket{0}_b \xrightarrow{\text{QRAM}} \sum_{j=1}^N \alpha_j \ket{j}_a \ket{D_j}_b$,
where $N=2^n$ is the memory size, and $\{\ket{j}_a\} = \{\ket{00...0},...,\ket{11...1}\}$ are the basis states of the $n$ address qubits $\opa_k,k=0,...,n-1$.

\begin{figure}[t]
	\centering
	\includegraphics[width=0.45\textwidth]{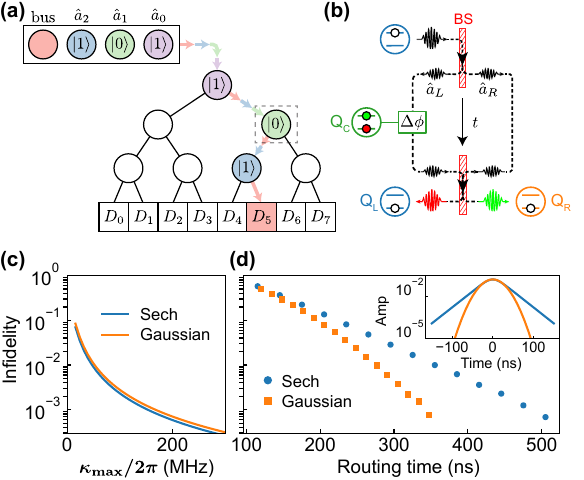}
	\caption{(a) Schematic of a bucket-brigade QRAM, illustrated with one particular query path. (b) The transmon-controlled phonon router with routing steps in the time domain. (c) Infidelity of the phonon routing as a function of maximum coupling strength $\kappa_{\text{max}}$. (d) Time domain simulated infidelity versus phonon routing time for hyperbolic secant and Gaussian pulse with fixed maximum coupling $\kappa_{\text{max}}=2\pi\times 200$ MHz. The inset shows hyperbolic secant and Gaussian pulse function used in (c) and (d) with the same FWHM = 50 ns.}
	\label{fig1}
\end{figure}

Among various QRAM architectures~\cite{giovannetti2008a,giovannetti2008,park2019,matteo2020,paler2020,niu2022,jaques2023,phalak2023,xu2023,hann2021a}, the bucket-brigade design~\cite{giovannetti2008a} is known for its efficiency and noise resilience~\cite{hann2021}, offering logarithmic query time scaling and polylogarithmic query infidelity scaling with respect to the memory size~\cite{hann2021}.
Schematically, a bucket-brigade QRAM consists of a binary tree of router nodes (Fig.~\ref{fig1}(a)), with address and data qubits sequentially routed into the tree from the root node.
Conditional routing is the fundamental operation in a bucket-brigade QRAM. At the $k$th level of the tree, the address qubit $\opa_k$ acts as the control, directing subsequent incident qubits left or right based on the control qubit's state.

The experimental realization of a bucket-brigade QRAM hinges on implementing scalable conditional routing.
Several candidate platforms have been proposed for QRAM, such as neutral atoms~\cite{giovannetti2008,hong2012,moiseev2016}, superconducting circuits~\cite{weiss2024}, photonic~\cite{chen2021} and phononic systems~\cite{hann2019}, in which the conditional routing relies on light-atom~\cite{giovannetti2008,hong2012,moiseev2016} or light-spin~\cite{chen2021} couplings, or nonlinear interactions mediated by transmon qubits~\cite{hann2019,weiss2024}.
So far a deterministic quantum router has only been demonstrated in superconducting circuits~\cite{naik2017,gao2019,wang2021}.
Although a random access quantum memory with classical addressing has been experimentally achieved~\cite{naik2017}, QRAM utilizing quantum addressing has yet to be realized.
A major challenge hindering QRAM development is loss errors due to energy relaxation.
Near-term experiments may address this issue through error detection combined with dual-rail encoding, although the required hardware cost is doubled~\cite{weiss2024}.

\begin{figure*}[t]
	\centering
	\includegraphics[width=0.7\textwidth]{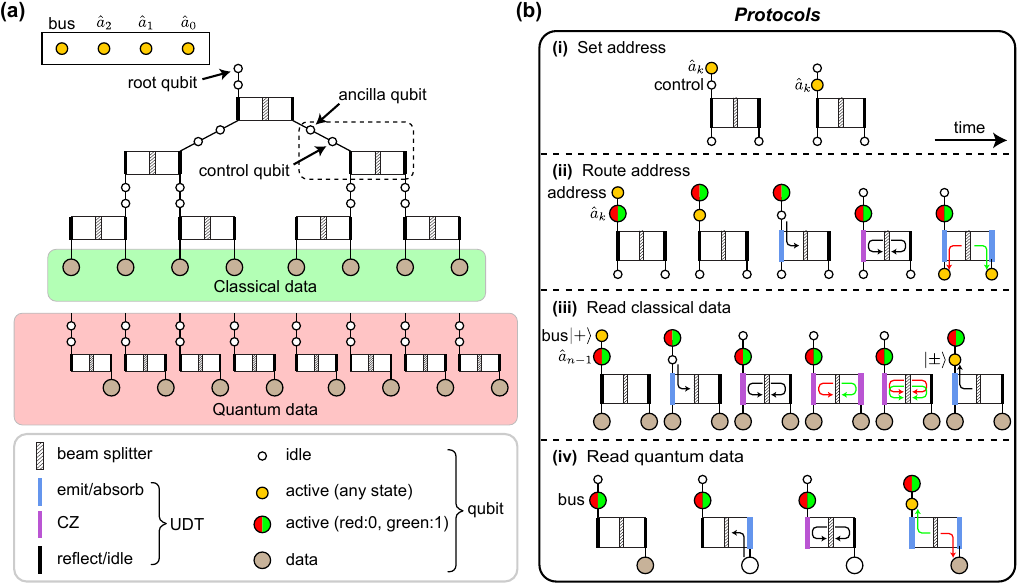}
	\caption{(a) Schematic hardware architecture to implement a bucket-brigade QRAM with transmon-controlled phonon routers. Symbols used for different hardware elements and operations are shown at the bottom. (b) Protocols for realizing the elementary operations required for a QRAM query.}
	\label{fig2}
\end{figure*}

We propose a hardware implementation of the generic bucket-brigade QRAM based on a hybrid platform of transmon qubits and surface acoustic wave (SAW) phonons.
The core of our design is a transmon-controlled phonon router, where itinerant phonons in a SAW waveguide are routed by a control transmon qubit.
Our approach offers several advantages over alternative proposals, including rapid query time, compact size, and the absence of frequency crowding.
We further introduce a hybrid dual-rail encoding scheme that enables loss error detection without additional hardware.
Due to the fast routing capability of the phonon router, our QRAM design achieves $\sim$kilohertz heralding rates for 100 memory cells, with realistic $T_1$ times of 100~{\textmu}s for the transmon and 2~{\textmu}s for the itinerant phonons.

\emph{Conditional phonon routing.}\textemdash
The phonon router (Fig.~\ref{fig1}(b)) consists of a SAW waveguide with a centered 50/50 phonon beam splitter (BS) and transmon qubits coupled to the ends of the waveguide via unidirectional transducers (UDT), a setup which has recently been demonstrated experimentally \cite{qiao2023}.
The qubits are coupled to the UDTs via tunable couplers \cite{chen2014}, allowing the qubits to emit into and absorb from a selected phonon mode of the SAW waveguide, with full control over the phonon temporal envelope \cite{bienfait2019}.
Furthermore, a reflective controlled-Z (CZ) gate between the transmon and an incident phonon can be performed by setting the transmon $e \leftrightarrow f$ transition resonant with the incoming phonon. Upon reflection, the phonon acquires an additional $\pi$ phase shift for the transmon in $\ket{e}$, with no phase shift for the transmon in $\ket{g}$~\cite{qiao}. We note this protocol relies on the finite bandwidth of the transducer, such that the $e \leftrightarrow g$ transition is not coupled to a SAW mode during the $e \leftrightarrow f$ resonant scattering process \cite{bienfait2019, bienfait2020}.

The conditional phonon routing requires one CZ gate sandwiched by two beam splitter interactions (Fig.~\ref{fig1}(b)). Similar to a Mach–Zehnder interferometer, the output phonon direction is determined by the relative phase shift between the two paths, the phase set by the transmon state.
More specifically, we first excite a left phonon mode $\opa_L$ and the beam splitter scatters the left and right modes into $\opa_L \rightarrow \frac{1}{\sqrt{2}} (-\opa_L + \opa_R)$ and $\opa_R \rightarrow \frac{1}{\sqrt{2}} (\opa_L + \opa_R)$.
The control transmon then applies a CZ gate to the left mode $\opa_L$, following which the left and right modes interfere again at the beam splitter, leading to an output $\opa_L$ or $\opa_R$ for the transmon in $\ket{g}$ or $\ket{e}$, respectively, which completes the conditional phonon routing (Fig.~\ref{fig1}(b)).

The performance of the phonon router is dependent on the highest achievable coupling rate $\kappa_{\text{max}}$ between the transmon qubit and the SAW waveguide. Physically, the maximal coupling $\kappa_{\text{max}}$ is constrained by the bandwidth of the transducer~\cite{bienfait2019, bienfait2020}, which has to be smaller than the transmon anharmonicity to protect the $e \leftrightarrow g$ transition from the $e \leftrightarrow f$ resonant scattering process. At a given maximum coupling rate, there is a trade-off in choosing the length of the emitted phonon wave packets: A shorter wave packet supports faster routing, reducing loss in the system. However, the increased phonon bandwidth concomitantly reduces the fidelity of the CZ gate, due to phonon wavepacket distortion on reflection from the UDT.
In Fig.~\ref{fig1}(c), we plot the routing infidelity as a function of $\kappa_{\text{max}}$ for hyperbolic secant and Gaussian phonon wavepacket shapes. We choose the full width at half maximum (FWHM) in either case to be 50 ns, close to what has been demonstrated experimentally \cite{bienfait2019,qiao2023}.
The routing time is non-zero, where the tail of the wavepacket contributes to the infidelity. We show in Fig.~\ref{fig1}(d) that Gaussian wave packets have faster-decaying tails and thus lower infidelity for a given routing time. We choose a routing time of 350~ns with a predicted routing infidelity of $10^{-3}$, comparable to the infidelity of single-qubit gates. Details about the infidelity calculations and time domain evolution simulations are in Appendix~\ref{sec:cswap}.

\emph{QRAM implementation.}\textemdash
Multiple phonon routers can be connected in a tree-like architecture to implement the bucket-brigade QRAM, as illustrated in Fig.~\ref{fig2}(a).
Each node (dashed rectangle in Fig.~\ref{fig2}(a)) comprises a control qubit for routing phonons in a SAW waveguide and an ancilla qubit for temporary qubit storage. We name the ancilla of the top node as the root qubit.
Prior to a query (Fig.~\ref{fig2}(a)), the address register holds active qubits (yellow filled circles) that carry quantum information, while the control and ancilla qubits in the QRAM remain idle (empty circles) in their ground states.
At the bottom of the tree, the routers interface with the data register, which is designed differently depending on whether the data is classical or quantum.

A complete QRAM query consists of three steps: routing addresses in, reading data, and routing addresses out.
The process begins by setting $\opa_0$ as the control qubit at the highest level of the QRAM tree. Then, sequentially from $k=1$ to $k=n-1$, we route the address qubit $\opa_k$ through the tree, conditioned on the states of $\opa_0,...,\opa_{k-1}$, setting $\opa_k$ as the control qubit for level $k$.
Once all address qubits are set, a bus qubit is routed in to retrieve the data.
Finally, we route the address qubits in a time-reversed process, disentangling the address register from the QRAM.

The elementary operations of setting and routing addresses required for a QRAM query are implemented as shown in Fig.~\ref{fig2}(b).
Once each address qubit $\opa_k$ is routed and stored in the ancilla qubit at level $k$, we can swap these with the control qubits at the same level to set the data address (Fig.~\ref{fig2}(b)i).
To route an address qubit stored in the ancilla at level $k$,the address qubit is emitted as a phonon, routed by the control qubit and stored in the corresponding left or right ancilla qubit at level $k+1$(Fig.~\ref{fig2}(b)ii).

To read classical data (Fig.~\ref{fig2}(b)iii), the bus qubit is initialized in $\ket{+}$. After all conditional routing, the bus qubit, now a phonon at level $n-1$, reaches the data qubits that are queried. We set the data qubits to $\ket{0}$ or $\ket{1}$ for classical data of 0 or 1, and perform CZ gates between the data qubits and the bus phonon. As a result, the bus phonon becomes $\ket{+}$ or $\ket{-}$ carrying the classical bit information.

To read quantum data (Fig.~\ref{fig2}(b)iv), the bus qubit is initialized in $\ket{1}$ and routed to the control qubits in the quantum data register.
Depending on whether a data qubit is queried or not, its control qubit will be in either $\ket{1}$ or $\ket{0}$.
Next, all data qubits are emitted into phonons and routed by the control qubits.
If the control qubit is in $\ket{1}$, i.e., the data qubit is queried, the data phonon will be routed into the QRAM and retrieved afterward.
Conversely, if the control qubit is in $\ket{0}$, i.e., the data qubit is not queried, the data phonon will be routed back and restored in the original data qubit.

\begin{figure}
	\centering
	\includegraphics[width=0.4\textwidth]{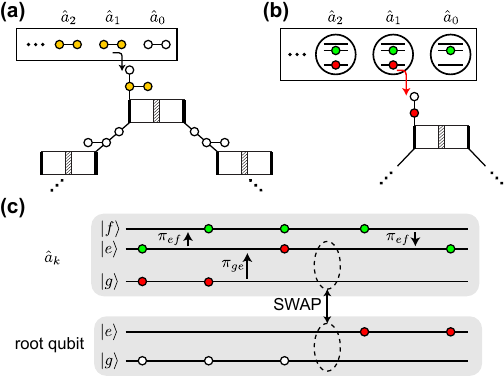}
	\caption{(a) The standard dual-rail encoding can be implemented by doubling the number of transmon qubits for the address, bus, and control qubits. (b) The hybrid dual-rail encoding does not require additional qubits, where only the ground state $\ket{g}$ of the address or bus qubit $\opa_k$ is released into the QRAM as an excitation. (c) Procedures for creating a dual-rail entangled pair between $\opa_k$ and the root qubit of the QRAM: $(\alpha \ket{g} + \beta \ket{e}) \ket{g} \xrightarrow{\pi_{ef}} (\alpha \ket{g} + \beta \ket{f}) \ket{g} \xrightarrow{\pi_{ge}} (\alpha \ket{e} + \beta \ket{f}) \ket{g} \xrightarrow{\text{SWAP}} \alpha \ket{g}\ket{e} + \beta \ket{f}\ket{g} \xrightarrow{\pi_{ef}} \alpha \ket{g}\ket{e} + \beta \ket{e}\ket{g}$. The red and green circles represent logical information of the address qubit.}
	\label{fig3}
\end{figure}

\emph{Error detection with dual-rail encodings.}\textemdash
Excitation loss due to transmon or phonon decay can be detected with dual-rail encoding~\cite{weiss2024}. The logical qubit is encoded in the one-excitation subspace $\{\ket{10},\ket{01}\}$ of two physical qubits, and the error state $\ket{00}$ is outside the logical subspace and thus is detectable.
In near-term experiments, such an error detection scheme may enable the demonstration of QRAM even in presence of dominant loss errors.

The standard dual-rail encoding leads to increased hardware complexity~\cite{weiss2024}. For our single-rail design (Fig.~\ref{fig2}(a)), we can implement dual-rail encoding by doubling the number of transmon qubits for the address, bus, and control qubits, while keeping the same number of ancilla qubits (Fig.~\ref{fig3}(a)).
Notably, phonon routing controlled by a dual-rail logical qubit can be achieved with phonon routing conditional on either of the physical qubits.
Since the two physical qubits are routed sequentially, the standard dual-rail encoding requires a longer query time (see Appendix~\ref{sec:query_time}).

For standard dual-rail encoding, we can initialize the QRAM in either vacuum states or the dual-rail subspace, resulting in two different error detection schemes~\cite{weiss2024}.
If the QRAM is initialized in vacuum states, error detection during a query would reveal which-way information. As a result, error detection can only be performed on the address and bus qubits at the end of a QRAM query.
This approach does not compromise error detection capability, as any loss error occurring during the query will be detectable at the end.
Alternatively, initializing the control qubits of all QRAM nodes in the dual-rail subspace allows real-time error detection during queries without revealing path information. However, this approach leads to exponentially more excitations, and thus an exponential reduction in the heralding rates (see Appendix~\ref{sec:heralding_rate}).

We propose a hybrid dual-rail encoding scheme that enables loss error detection with the single-rail design (Fig.~\ref{fig2}(a)), avoiding the additional hardware required for the standard dual-rail encoding.
The only difference from the single-rail case lies in how the address or bus qubits are released into the QRAM.
Instead of a full swap of the address or bus qubit into the root qubit of the QRAM, we first create a dual-rail entangled pair between them and then route half of the pair into the QRAM (Fig.~\ref{fig3}(b)).
More precisely, we entangle the address or bus qubit $\opa_k$ with the root qubit through the mapping:
\begin{equation}
    (\alpha \ket{0}_k + \beta \ket{1}_k) \ket{0}_{\text{root}} \rightarrow \alpha \ket{0}_k \ket{1}_{\text{root}} + \beta \ket{1}_k \ket{0}_{\text{root}} ,
\end{equation}
and route the root qubit into the QRAM, where $k=n$ denotes the bus qubit.
This mapping can be implemented by utilizing the second excited state $\ket{f}$ of the address or bus transmon qubit $\opa_k$ (Fig.~\ref{fig3}(c)).

For hybrid dual-rail encoding, the QRAM is initialized in vacuum states and error detection happens at the end of the query.
After routing out from the QRAM and reversing the procedures in Fig.~\ref{fig3}(c), the retrieved address and bus qubits should remain in the subspace of $\{\ket{g},\ket{e}\}$ if no loss has occurred.
On the other hand, the swapping and routing operations preserve the number of excitations and thus any decay during the query results in one or more address and bus qubits in $\ket{f}$, which is detectable.
Here we ignore the decay from $\ket{f}$ to $\ket{e}$ in Fig.~\ref{fig3}(c), as its contribution to the overall query infidelity is negligible due to the significantly shorter time occupying $\ket{f}$ compared to the routing time (see Appendix~\ref{sec:heralding_fidelity}).
The hybrid dual-rail encoding is applicable to other QRAM implementations~\cite{hann2019,weiss2024} without changes to the hardware architectures.

\begin{figure}
	\centering
	\includegraphics[width=0.4\textwidth]{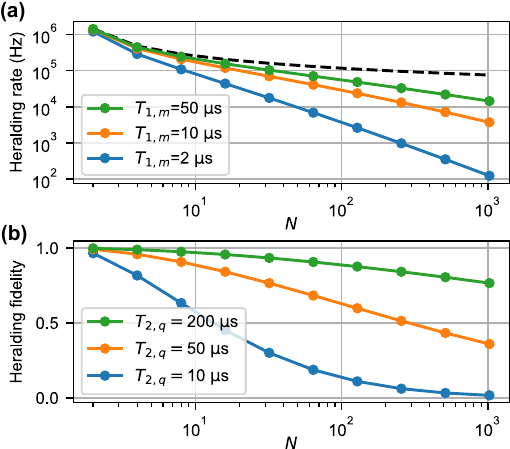}
	\caption{(a) Estimated heralding rates for different memory sizes $N=2^n$ and phonon lifetimes $T_{1,m}$, where the transmon lifetime is $T_{1,q}=100$~{\textmu}s. The black dashed line corresponds to $T_{1,q}$ and $T_{1,m}$ both being infinite. (b) Lower bound on the heralding fidelity due to transmon dephasing errors, where phonon is assumed to be dephasing-free.}
	\label{fig4}
\end{figure}

\emph{Heralding rate and fidelity estimations.}\textemdash
Here we estimate the heralding rate for the hybrid dual-rail encoding. Since any decay during the QRAM query is detectable, the heralding rate is given by $P(\text{no error})/T$, where $T$ is the total query time and $P(\text{no error})$ is the probability of no decay occurring during the query.
Assuming each routing step takes time $t$, the total query time can be counted as $T=2(2n-1)t$ (see Appendix~\ref{sec:query_time}).
In the hybrid dual-rail encoding, each address and bus qubit contributes one excitation. The success probability can be estimated from the duration each excitation undergoes transmon decay or phonon decay (see Appendix~\ref{sec:heralding_rate}).
For $T_{1,q}=T_{1,m} \equiv T_1$ where $T_{1,q}$ and $T_{1,m}$ are the energy relaxation times of the transmon and phonon respectively, the success probability is $P(\text{no error}) = \exp \left( -(n+1) T/T_1 \right)$.
A more general estimation is provided in Appendix~\ref{sec:heralding_rate}.

We plot the heralding rates for different memory sizes $N=2^n$ and phonon lifetimes $T_{1,m}$ in Fig.~\ref{fig4}(a), with a transmon lifetime of $T_{1,q}=100$~{\textmu}s and a phonon routing time of $t=350$~ns. Due to the fast phonon routing, we can achieve a few kilohertz heralding rates for 100 data qubits using current device parameters \cite{qiao2023}.
The heralding rate decreases for larger memory sizes due to the higher decay probability and longer query times. With realistic improvement of phonon lifetime, the heralding rate can approach an upper bound without transmon or phonon decay, where the heralding rate is determined solely by the total query time (denoted by the back dashed line).

If energy relaxation were the only noise process, the query infidelity after heralding would be determined by routing errors. Wave packet distortion affects path interference (Fig.~\ref{fig1}(b)), leading to incorrect routing direction.
In hybrid dual-rail encoding, each address qubit $\opa_k,k=0,\ldots,n-1$ contributes exactly one excitation which gets routed 0 or $2k$ times during a QRAM query.
We can estimate the query infidelity from the total number of routing steps as $1-F \sim \epsilon n(n-1)$ with $\epsilon$ being the infidelity per routing step (Fig.~\ref{fig1}(c)).

Additional noise processes such as dephasing and thermal noise may further reduce the query fidelity since they cannot be detected by the dual-rail encoding.
Assuming the worst-case scenario where any dephasing error results in zero fidelity (see Appendix~\ref{sec:heralding_fidelity}), we calculate a lower bound on query fidelity for different memory sizes and transmon dephasing times $T_{2,q}$ in Fig.~\ref{fig4}(b), which scales as $1-F \sim 2n^2 t/T_{2,q}$ for small errors.
Since thermal excitations may propagate errors within the QRAM, its infidelity scales as $1-F \lesssim 4 \bar{n}_{\text{th}} n(n+1)T/T_1 \sim 16 \bar{n}_{\text{th}} n^3 t/T_1$~\cite{hann2021}, where $\bar{n}_{\text{th}}$ is the average thermal occupation of the environment. Since $\bar{n}_{\text{th}}$ is less than 1\% at millikelvin temperatures and gigahertz frequencies \cite{satzinger2018}, for near-term shallow depth QRAM where $n \bar{n}_{\text{th}} \ll 1$, we expect the infidelity to be dominated by dephasing errors.

\emph{Discussion.}\textemdash
We have proposed a QRAM implementation based on a hybrid platform of transmon-controlled phonon routers. We have also introduced the hybrid dual-rail encoding which detects loss errors without any additional hardware, and have estimated the heralding rate and fidelity. Our general protocol applies to other itinerant phonon or photon-based platforms \cite{zivari2021, chen2021,kuzyk2018,lemonde2018}.

The hardware platform we consider offers several advantages.
Using itinerant phonons in our design not only provides a more compact solution compared to 3D microwave cavities~\cite{weiss2024}, but also avoids frequency crowding in multimode systems~\cite{hann2019,naik2017}.
Additionally, our phonon router relies on linear phonon scattering which is fast and does not require ancilla qubits to facilitate the nonlinear interaction.

In future work, several optimizations of our proposed QRAM implementation are within reach, such as mitigating dephasing noise through rapid echo techniques in dual-rail encodings and enabling direct phonon transfer between SAW waveguides without ancilla qubits.
While error detection using dual-rail encodings is effective for near-term experimental demonstrations of QRAM, exploring alternative qubit codes for both transmons and phonons is crucial for implementing full quantum error correction and ultimately achieving fault tolerance in QRAM~\cite{hann2021,matteo2020}.
Moreover, the hybrid transmon-phonon platform holds promise for broader applications in quantum computation, including cluster state generation~\cite{zhan2020} and establishing non-local qubit connectivity.

\begin{acknowledgments}
    
    We acknowledge support from the ARO(W911NF-23-1-0077), ARO MURI (W911NF-21-1-0325), the Air Force Office of Scientific Research (AFOSR grant FA9550-20-1-0270), AFOSR MURI (FA9550-19-1-0399, FA9550-21-1-0209, FA9550-23-1-0338), DARPA (HR0011-24-9-0359, HR0011-24-9-0361), NSF (OMA-1936118, ERC-1941583, OMA-2137642, OSI-2326767, CCF-2312755), NTT Research, Packard Foundation (2020-71479), and the Marshall and Arlene Bennett Family Research Program. This work was partially supported by the Defense Advanced Research Projects Agency (DARPA) under Agreement No. HR00112490364, by UChicago's MRSEC (NSF award DMR-2011854) and by the NSF QLCI for HQAN (NSF award 2016136). This material is based upon work supported by the U.S. Department of Energy, Office of Science, National Quantum Information Science Research Centers and Advanced Scientific Computing Research (ASCR) program under contract number DE-AC02-06CH11357 as part of the InterQnet quantum networking project, as well as the U.S. Department of Energy Office of Science National Quantum Information Science Research Centers.
\end{acknowledgments}

\appendix

\section{Qubit controlled phonon routing}
\label{sec:cswap}

\subsection{Time domain simulation and infidelity}
\begin{figure*}
	\centering
	\includegraphics[width=0.8\textwidth]{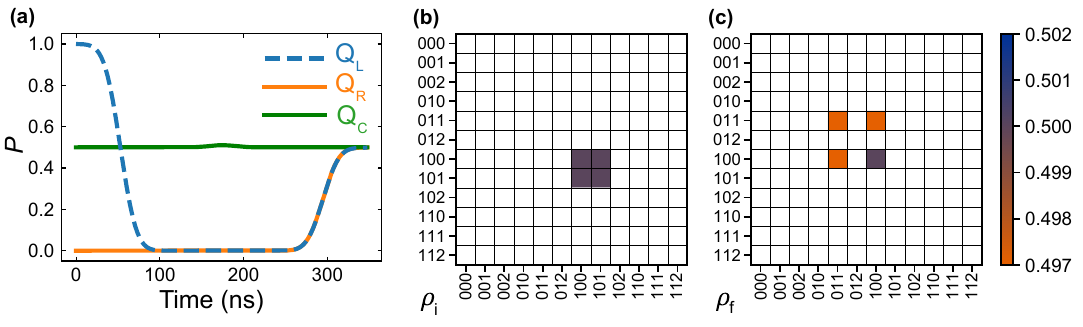}
	\caption{(a) Time domain simulation of phonon routing with the states of Q$_L$, Q$_R$ and Q$_C$ being initialized to $\ket{1}$, $\ket{0}$ and $(\ket{0}+\ket{1})\sqrt{2}$ respectively. We use parameters in main text $\kappa_{\bf max}=2\pi\times 200$ MHz and Gaussian FWHM = 50 ns. The initial ($\rho_i$) and final density matrix ($\rho_f$) at time $t_i=0$ ns and $t_f=350$ ns are plotted in (b) and (c). We only show non-zero density matrix elements with real values close to 0.5 that contribute to the fidelity.}
	\label{figS1}
\end{figure*}
Here we show time domain simulation \cite{kiilerich2019} of the process shown in Fig.~\ref{fig1}(b) as an example of a conditional phonon routing. We start with the left transmon qubit excited, right transmon qubit in its ground state, and the control qubit in superposition state $\ket{\phi_c}=(\ket{0}+\ket{1})/\sqrt{2}$. So the initial 3-qubit state is 
\begin{equation}
\ket{\psi_i}=(\ket{100}+\ket{101})/\sqrt{2},
\end{equation}
with state label in the order left, right and control qubit. As shown in Fig.~\ref{figS1}(a), left qubit $Q_L$ releases a single phonon into left phonon mode $\opa_L$ with Gaussian mode profile $\sqrt[4]{\frac{2 \kappa^2}{\pi}} e^{-(\kappa t)^2}$, split by the BS and scattered by the control qubit $Q_C$ on one side. After another interference between the scattered and unscattered phonon mode at BS, $\opa_L$ and $\opa_R$ are captured by the $Q_L$ and $Q_R$ respectively with final state $\rho_f$. We plot each qubit population evolved with time. The population from $Q_L$ is transferred to an equal superposition of $Q_L$ and $Q_R$ as expected after a CSWAP gate. $Q_C$ population is maintained at 0.5 with a tiny pump in the middle indicating the scattering process without significantly populating the qubit. We also show the initial density matrix $\rho_i=\ket{\psi_i}\bra{\psi_i}$ and final qubit density matrix $\rho_f$ after CSWAP gate in Fig.~\ref{figS1}(b) and (c). The simulated fidelity of $\rho_f$ to the ideal final density matrix $\rho_\text{ideal}=\ket{\psi_\text{ideal}}\bra{\psi_\text{ideal}}$ is 
\begin{equation}
\mathcal{F} = \Tr(\rho_f\cdot\rho_\text{ideal})=0.9992,
\end{equation}
where $\ket{\psi_\text{ideal}}=(\ket{100}+\ket{011})/\sqrt{2}$ given only $Q_L$ and $Q_R$ swap happen when $Q_C$ is at $\ket{1}$. We use 2-level qubits for $Q_L$ and $Q_R$ and 3 levels for $Q_C$ as $e \leftrightarrow f$ transition is used for scatter. We note that the influence of $T_1$ and $T_2$ has been included in the heralding rate and fidelity analysis so this simulation assumes no decay or dephasing.

\subsection{Infidelity due to distortion}
Here we give an analytical expression to the distortion-induced infidelity, where an infinite routing time is assumed to ignore any infidelity from the wave packet tails.
To start, let us denote a single phonon state in the SAW waveguide with frequency profile $u(\omega)$ as $\ket{u}$. Upon reflection on a transmon qubit with linewidth $\kappa_{\text{max}}$, the single phonon state gets distorted and becomes $\ket{v}$, where
\begin{equation}
    \label{eq:distortion}
    v(\omega) = u(\omega) \frac{i\omega + \kappa_{\text{max}}/2}{i\omega - \kappa_{\text{max}}/2} ,
\end{equation}
which can be derived from the input-output relation.

For the simulation above, after releasing the phonon from $Q_L$ and the beam splitter, the state of the system is
\begin{equation}
    \ket{\phi_c} \otimes \frac{1}{\sqrt{2}} (-\ket{u}_L \ket{0}_R + \ket{0}_L \ket{u}_R) .
\end{equation}
After reflection on the control qubit, the system state becomes
\begin{equation}
    \begin{split}
        \ket{\Psi} =& \frac{1}{2} \ket{0}_c \otimes (-\ket{u}_L \ket{0}_R + \ket{0}_L \ket{u}_R) \\
        & + \frac{1}{2} \ket{1}_c \otimes (-\ket{v}_L \ket{0}_R + \ket{0}_L \ket{u}_R) .
    \end{split}
\end{equation}
The ideal final state is
\begin{equation}
    \frac{1}{\sqrt{2}} (\ket{0}_c \ket{u}_L \ket{0}_R + \ket{1}_c \ket{0}_L \ket{u}_R) ,
\end{equation}
which is
\begin{equation}
    \begin{split}
        \ket{\Psi_\text{ideal}} =& \frac{1}{2} \ket{0}_c \otimes (-\ket{u}_L \ket{0}_R + \ket{0}_L \ket{u}_R) \\
        & + \frac{1}{2} \ket{1}_c \otimes (\ket{u}_L \ket{0}_R + \ket{0}_L \ket{u}_R) .
    \end{split}
\end{equation}
before the second beam splitter.

The fidelity is given by
\begin{equation}
    \begin{split}
        \mathcal{F} =& |\inp{\Psi_\text{ideal}}{\Psi}|^2 = \left| \frac{3}{4} + \frac{1}{4} \inp{u}{-v} \right|^2 \\
        =& \left| \frac{3}{4} + \frac{1}{4} \int_{-\infty}^{\infty} |u(\omega)|^2 \frac{i\omega + \kappa_{\text{max}}/2}{-i\omega + \kappa_{\text{max}}/2} \text{d}\omega \right|^2 .
    \end{split}
\end{equation}
For symmetric wave packet where $|u(\omega)|=|u(-\omega)|$, we have
\begin{equation}
    \begin{split}
        \mathcal{F} =& \left( \frac{3}{4} + \frac{1}{4} \int_{-\infty}^{\infty} |u(\omega)|^2 \frac{\kappa_{\text{max}}^2/4 - \omega^2}{\kappa_{\text{max}}^2/4 + \omega^2} \text{d}\omega \right)^2 \\
        =& \left( 1 - 2\int_{-\infty}^{\infty} |u(\omega)|^2 \frac{\omega^2}{\kappa_{\text{max}}^2 + 4\omega^2} \text{d}\omega \right)^2 ,
    \end{split}
\end{equation}
which leads to the results in Fig.~\ref{fig1}(c).

Physically, the maximal coupling $\kappa_{\text{max}}$ is constrained by the bandwidth of the transducer~\cite{bienfait2019, bienfait2020}, which has to be smaller than the transmon anharmonicity to protect the $e \leftrightarrow g$ transition from the $e \leftrightarrow f$ resonant scattering process.
Meanwhile, as long as the bandwidth of the phonon wave packet is much smaller than the transducer bandwidth, Eq.~(\ref{eq:distortion}) is a good approximation.

\section{Query time estimation}
\label{sec:query_time}
\subsection{Hybrid dual-rail encoding}
Here we estimate the total QRAM query time, including routing in and routing out stages. We ignore the time costs of transmon operations since the total time is dominated by phonon routing.
We assume each routing step takes time $t$, and the time is counted from a transmon releasing into the phonon waveguide to a transmon re-absorbing the phonon.

The routing time of the address qubits $\opa_k,k=0,...,n-1$ and the bus qubit $\opa_{k=n}$ is $2kt$, and the total QRAM query time is $T=2(2n-1)t$ where $n$ is the number of address qubits.
Notice that the total time $T$ scales linearly with $n$ instead of quadratically since the routing at different QRAM levels can be performed in parallel.
Fig.~\ref{fig5}(a) shows the routing schedule of each address and bus qubit for $n=4$.

\subsection{Standard dual-rail encoding}
\begin{figure}
	\centering
	\includegraphics[width=0.48\textwidth]{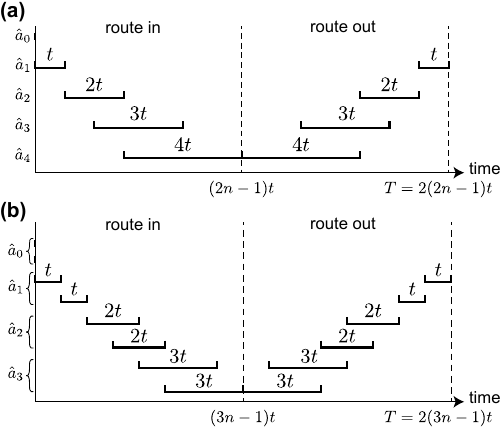}
	\caption{Routing schedule for (a) hybrid dual-rail encoding and (b) standard dual-rail encoding.}
	\label{fig5}
\end{figure}

For standard dual-rail encoding, both physical qubits in the encoding need to be routed, leading to a increase of the total query time. With the routing schedule in Fig.~\ref{fig5}(b), the total query time is $T=2(3n-1)t$.

\section{Heralding rate estimation}
\label{sec:heralding_rate}
In the section, we estimate the heralding rates for three different scenarios: hybrid dual-rail encoding, standard dual-rail encoding with QRAM initialized in vacuum states, standard dual-rail encoding with QRAM initialized in the logical subspace.

\subsection{Hybrid dual-rail encoding}
In hybrid dual-rail encoding, each address or bus qubit contributes exactly one excitation. For qubit $\opa_k,k=0,...,n$, the routing steps take time $2kt$ where the excitation exists either in a transmon or in a phonon waveguide throughout the QRAM operation. For the rest of time $T-2kt$ where $T=2(2n-1)t$, the excitation only exists in a transmon. Therefore, the average success probability with no excitation loss error is
\begin{equation}
    \label{eq:heralding_rate_estimation}
    \begin{split}
        & P(\text{no error}) \\
        =& \prod_{k=0}^n \frac{1}{2} \left( e^{-2kt/T_m} + e^{-2kt/T_q} \right) e^{-(T - 2kt)/T_q} \\
        =& \exp \left( -(n+1) \frac{T}{T_q} \right) \prod_{k=0}^n \frac{1}{2} \left[ 1 + \exp \left( 2kt (\frac{1}{T_q} - \frac{1}{T_m}) \right) \right] ,
    \end{split}
\end{equation}
where $T_q$ and $T_m$ are the $T_1$ lifetime of the transmon and phonon.
The average heralding rate is thus $P(\text{no error})/T$.

We can bound the success probability by $P_{\text{min}} \leq P(\text{no error}) \leq P_{\text{max}}$, where
\begin{equation}
    \begin{split}
        P_{\text{min}} =& \exp \left( - (n+1) \left( \frac{T}{T_q} - \frac{nt}{T_q} + \frac{nt}{\min (T_q,T_m)} \right) \right) \\
        P_{\text{max}} =& \exp \left( - (n+1) \left( \frac{T}{T_q} - \frac{nt}{T_q} + \frac{nt}{\max (T_q,T_m)} \right) \right) .
    \end{split}
\end{equation}
When $T_m=T_q$, we have $P(\text{no error}) = \exp (-(n+1)T/T_q)$.

\subsection{Standard dual-rail encoding with vacuum states initialization}
When the routers are initialized in vacuum states, all excitations come from the address and bus qubits and each address and bus qubit contributes one excitation. Each excitation exists in a phonon mode for a time $2kt$ and otherwise resides in a transmon. Similar to Eq.~(\ref{eq:heralding_rate_estimation}), the success probability is
\begin{equation}
    \begin{split}
        P(\text{no error}) =& \prod_{k=0}^n e^{-2kt/T_m} e^{-(T - 2kt)/T_q} \\
        =& \exp \left( - (n+1) \left( \frac{T}{T_q} - \frac{nt}{T_q} + \frac{nt}{T_m} \right) \right) ,
    \end{split}
\end{equation}
where $T=2(3n-1)t$.

\subsection{Standard dual-rail encoding with logical subspace initialization}
When the routers are initialized in the dual-rail subspace, each router also contributes one excitation and the total number of excitations in the QRAM scales as $2^n$. The advantage compared to the vacuum states initialization is that we can perform real-time error detection to all routers without leaking any which-way information.
The disadvantage, however, is that the success probability is much lower due to the exponential increase in the number of excitations.
Even if phonon loss is ignored, the success probability due to the transmon decay scales as
\begin{equation}
    P(\text{no error}) \sim \exp \left( - 2^n \frac{T}{T_q} \right) ,
\end{equation}
where the total error rate is exponentially high in $n$.

\section{Heralding fidelity estimation}
\label{sec:heralding_fidelity}
If photon loss is the only error in the system, the dual-rail encoding can detect all photon loss events and the heralding fidelity would be 1. However, other noise processes may exist which reduce the fidelity after heralding. Here we estimate the heralding fidelity due to dephasing error.

For a single qubit undergoing a continuous dephasing process for time $t$, the Kraus operators of the resulting dephasing channel are $\{ \sqrt{1-\frac{p}{2}} I, \sqrt{\frac{p}{2}} Z \}$ with $p = 1 - e^{-t/T_2}$.
Therefore, the probability of no dephasing error for a single qubit is
\begin{equation}
    p(t,T_2) = \frac{1 + e^{-t/T_2}}{2} .
\end{equation}

We can estimate the probability of no dephasing error during a QRAM query.
Similar to photon loss, dephasing errors only impact the excitations in the system and do not impact the idle routers in vacuum states.
Therefore, the no error probability is
\begin{equation}
    \begin{split}
        & P(\text{no error}) \\
        =& \prod_{k=0}^n \frac{p(2kt,T_{2,m}) + p(2kt,T_{2,q})}{2} p(T-2kt,T_{2,q}) ,
    \end{split}
\end{equation}
where $T_{2,q}$ and $T_{2,m}$ are the $T_2$ times of the transmon and phonon.
For small errors and $T_{2,m} \rightarrow \infty$, we have $1-P(\text{no error}) \sim (n+1) T/2 T_{2,q} \sim 2n^2t / T_{2,q}$.

If we assume that the fidelity drops to zero under any dephasing error, the query fidelity is lower bounded by $P(\text{no error})$. More generally, the fidelity due to dephasing error is path dependent. For example, if we query only one path without any superpositions, dephasing error does not have any impact.
Furthermore, we can circumvent dephasing errors with techniques such as dynamical decoupling.
Longer phonon wave packets increase dephasing errors but reduce distortion-induced infidelity, suggesting the existence of an optimal wave packet length in practice.

Throughout the paper, we have neglected the decay from $\ket{f}$ to $\ket{e}$ during the preparation of the dual-rail entangled pair between the address and root qubits (Fig.~\ref{fig3}(c)). This is because such decay processes occur only at the first level of the QRAM and have a significantly shorter duration than the routing time. The probability of no decay from $\ket{f}$ to $\ket{e}$ approximately scales as $\exp (-n t_f / T_{1,q})$, where $t_f \ll t$ denotes the duration for which $\ket{f}$ is occupied and can be shorter than 50~ns~\cite{bianchetti2010,chen2014}.
Therefore, the contributions from $\ket{f}$ to $\ket{e}$ decay is negligible compared to the fidelity drop due to dephasing which scales as $\exp (-2n^2 t / T_{2,q})$.

\bibliography{SAW_based_QRAM}

\begin{thebibliography}{41}%
\makeatletter
\providecommand \@ifxundefined [1]{%
 \@ifx{#1\undefined}
}%
\providecommand \@ifnum [1]{%
 \ifnum #1\expandafter \@firstoftwo
 \else \expandafter \@secondoftwo
 \fi
}%
\providecommand \@ifx [1]{%
 \ifx #1\expandafter \@firstoftwo
 \else \expandafter \@secondoftwo
 \fi
}%
\providecommand \natexlab [1]{#1}%
\providecommand \enquote  [1]{``#1''}%
\providecommand \bibnamefont  [1]{#1}%
\providecommand \bibfnamefont [1]{#1}%
\providecommand \citenamefont [1]{#1}%
\providecommand \href@noop [0]{\@secondoftwo}%
\providecommand \href [0]{\begingroup \@sanitize@url \@href}%
\providecommand \@href[1]{\@@startlink{#1}\@@href}%
\providecommand \@@href[1]{\endgroup#1\@@endlink}%
\providecommand \@sanitize@url [0]{\catcode `\\12\catcode `\$12\catcode
  `\&12\catcode `\#12\catcode `\^12\catcode `\_12\catcode `\%12\relax}%
\providecommand \@@startlink[1]{}%
\providecommand \@@endlink[0]{}%
\providecommand \url  [0]{\begingroup\@sanitize@url \@url }%
\providecommand \@url [1]{\endgroup\@href {#1}{\urlprefix }}%
\providecommand \urlprefix  [0]{URL }%
\providecommand \Eprint [0]{\href }%
\providecommand \doibase [0]{https://doi.org/}%
\providecommand \selectlanguage [0]{\@gobble}%
\providecommand \bibinfo  [0]{\@secondoftwo}%
\providecommand \bibfield  [0]{\@secondoftwo}%
\providecommand \translation [1]{[#1]}%
\providecommand \BibitemOpen [0]{}%
\providecommand \bibitemStop [0]{}%
\providecommand \bibitemNoStop [0]{.\EOS\space}%
\providecommand \EOS [0]{\spacefactor3000\relax}%
\providecommand \BibitemShut  [1]{\csname bibitem#1\endcsname}%
\let\auto@bib@innerbib\@empty
\bibitem [{\citenamefont {Giovannetti}\ \emph
  {et~al.}(2008{\natexlab{a}})\citenamefont {Giovannetti}, \citenamefont
  {Lloyd},\ and\ \citenamefont {Maccone}}]{giovannetti2008a}%
  \BibitemOpen
  \bibfield  {author} {\bibinfo {author} {\bibfnamefont {V.}~\bibnamefont
  {Giovannetti}}, \bibinfo {author} {\bibfnamefont {S.}~\bibnamefont {Lloyd}},\
  and\ \bibinfo {author} {\bibfnamefont {L.}~\bibnamefont {Maccone}},\
  }\bibfield  {title} {\bibinfo {title} {Quantum {{Random Access Memory}}},\
  }\href {https://doi.org/10.1103/PhysRevLett.100.160501} {\bibfield  {journal}
  {\bibinfo  {journal} {Physical Review Letters}\ }\textbf {\bibinfo {volume}
  {100}},\ \bibinfo {pages} {160501} (\bibinfo {year}
  {2008}{\natexlab{a}})}\BibitemShut {NoStop}%
\bibitem [{\citenamefont {Grover}(1996)}]{grover1996}%
  \BibitemOpen
  \bibfield  {author} {\bibinfo {author} {\bibfnamefont {L.~K.}\ \bibnamefont
  {Grover}},\ }\bibfield  {title} {\bibinfo {title} {A fast quantum mechanical
  algorithm for database search},\ }in\ \href
  {https://doi.org/10.1145/237814.237866} {\emph {\bibinfo {booktitle}
  {Proceedings of the Twenty-Eighth Annual {{ACM}} Symposium on {{Theory}} of
  {{Computing}}}}},\ \bibinfo {series and number} {{{STOC}} '96}\ (\bibinfo
  {publisher} {Association for Computing Machinery},\ \bibinfo {address} {New
  York, NY, USA},\ \bibinfo {year} {1996})\ pp.\ \bibinfo {pages}
  {212--219}\BibitemShut {NoStop}%
\bibitem [{\citenamefont {Harrow}\ \emph {et~al.}(2009)\citenamefont {Harrow},
  \citenamefont {Hassidim},\ and\ \citenamefont {Lloyd}}]{harrow2009}%
  \BibitemOpen
  \bibfield  {author} {\bibinfo {author} {\bibfnamefont {A.~W.}\ \bibnamefont
  {Harrow}}, \bibinfo {author} {\bibfnamefont {A.}~\bibnamefont {Hassidim}},\
  and\ \bibinfo {author} {\bibfnamefont {S.}~\bibnamefont {Lloyd}},\ }\bibfield
   {title} {\bibinfo {title} {Quantum {{Algorithm}} for {{Linear Systems}} of
  {{Equations}}},\ }\href {https://doi.org/10.1103/PhysRevLett.103.150502}
  {\bibfield  {journal} {\bibinfo  {journal} {Physical Review Letters}\
  }\textbf {\bibinfo {volume} {103}},\ \bibinfo {pages} {150502} (\bibinfo
  {year} {2009})}\BibitemShut {NoStop}%
\bibitem [{\citenamefont {Biamonte}\ \emph {et~al.}(2017)\citenamefont
  {Biamonte}, \citenamefont {Wittek}, \citenamefont {Pancotti}, \citenamefont
  {Rebentrost}, \citenamefont {Wiebe},\ and\ \citenamefont
  {Lloyd}}]{biamonte2017}%
  \BibitemOpen
  \bibfield  {author} {\bibinfo {author} {\bibfnamefont {J.}~\bibnamefont
  {Biamonte}}, \bibinfo {author} {\bibfnamefont {P.}~\bibnamefont {Wittek}},
  \bibinfo {author} {\bibfnamefont {N.}~\bibnamefont {Pancotti}}, \bibinfo
  {author} {\bibfnamefont {P.}~\bibnamefont {Rebentrost}}, \bibinfo {author}
  {\bibfnamefont {N.}~\bibnamefont {Wiebe}},\ and\ \bibinfo {author}
  {\bibfnamefont {S.}~\bibnamefont {Lloyd}},\ }\bibfield  {title} {\bibinfo
  {title} {Quantum machine learning},\ }\href
  {https://doi.org/10.1038/nature23474} {\bibfield  {journal} {\bibinfo
  {journal} {Nature}\ }\textbf {\bibinfo {volume} {549}},\ \bibinfo {pages}
  {195} (\bibinfo {year} {2017})}\BibitemShut {NoStop}%
\bibitem [{\citenamefont {Ciliberto}\ \emph {et~al.}(2018)\citenamefont
  {Ciliberto}, \citenamefont {Herbster}, \citenamefont {Ialongo}, \citenamefont
  {Pontil}, \citenamefont {Rocchetto}, \citenamefont {Severini},\ and\
  \citenamefont {Wossnig}}]{ciliberto2018}%
  \BibitemOpen
  \bibfield  {author} {\bibinfo {author} {\bibfnamefont {C.}~\bibnamefont
  {Ciliberto}}, \bibinfo {author} {\bibfnamefont {M.}~\bibnamefont {Herbster}},
  \bibinfo {author} {\bibfnamefont {A.~D.}\ \bibnamefont {Ialongo}}, \bibinfo
  {author} {\bibfnamefont {M.}~\bibnamefont {Pontil}}, \bibinfo {author}
  {\bibfnamefont {A.}~\bibnamefont {Rocchetto}}, \bibinfo {author}
  {\bibfnamefont {S.}~\bibnamefont {Severini}},\ and\ \bibinfo {author}
  {\bibfnamefont {L.}~\bibnamefont {Wossnig}},\ }\bibfield  {title} {\bibinfo
  {title} {Quantum machine learning: A classical perspective},\ }\href
  {https://doi.org/10.1098/rspa.2017.0551} {\bibfield  {journal} {\bibinfo
  {journal} {Proceedings of the Royal Society A: Mathematical, Physical and
  Engineering Sciences}\ }\textbf {\bibinfo {volume} {474}},\ \bibinfo {pages}
  {20170551} (\bibinfo {year} {2018})}\BibitemShut {NoStop}%
\bibitem [{\citenamefont {Lloyd}\ \emph {et~al.}(2013)\citenamefont {Lloyd},
  \citenamefont {Mohseni},\ and\ \citenamefont {Rebentrost}}]{lloyd2013}%
  \BibitemOpen
  \bibfield  {author} {\bibinfo {author} {\bibfnamefont {S.}~\bibnamefont
  {Lloyd}}, \bibinfo {author} {\bibfnamefont {M.}~\bibnamefont {Mohseni}},\
  and\ \bibinfo {author} {\bibfnamefont {P.}~\bibnamefont {Rebentrost}},\
  }\href {https://doi.org/10.48550/arXiv.1307.0411} {\bibinfo {title} {Quantum
  algorithms for supervised and unsupervised machine learning}} (\bibinfo
  {year} {2013}),\ \Eprint {https://arxiv.org/abs/1307.0411} {arXiv:1307.0411
  [quant-ph]} \BibitemShut {NoStop}%
\bibitem [{\citenamefont {Kerenidis}\ and\ \citenamefont
  {Prakash}(2017)}]{kerenidis2017}%
  \BibitemOpen
  \bibfield  {author} {\bibinfo {author} {\bibfnamefont {I.}~\bibnamefont
  {Kerenidis}}\ and\ \bibinfo {author} {\bibfnamefont {A.}~\bibnamefont
  {Prakash}},\ }\bibfield  {title} {\bibinfo {title} {Quantum {{Recommendation
  Systems}}},\ }in\ \href {https://doi.org/10.4230/LIPIcs.ITCS.2017.49} {\emph
  {\bibinfo {booktitle}
  {{{DROPS-IDN}}/v2/Document/10.4230/{{LIPIcs}}.{{ITCS}}.2017.49}}}\ (\bibinfo
  {publisher} {Schloss Dagstuhl -- Leibniz-Zentrum f{\"u}r Informatik},\
  \bibinfo {year} {2017})\BibitemShut {NoStop}%
\bibitem [{\citenamefont {Kerenidis}\ and\ \citenamefont
  {Prakash}(2020)}]{kerenidis2020}%
  \BibitemOpen
  \bibfield  {author} {\bibinfo {author} {\bibfnamefont {I.}~\bibnamefont
  {Kerenidis}}\ and\ \bibinfo {author} {\bibfnamefont {A.}~\bibnamefont
  {Prakash}},\ }\bibfield  {title} {\bibinfo {title} {Quantum gradient descent
  for linear systems and least squares},\ }\href
  {https://doi.org/10.1103/PhysRevA.101.022316} {\bibfield  {journal} {\bibinfo
   {journal} {Physical Review A}\ }\textbf {\bibinfo {volume} {101}},\ \bibinfo
  {pages} {022316} (\bibinfo {year} {2020})}\BibitemShut {NoStop}%
\bibitem [{\citenamefont {Cao}\ \emph {et~al.}(2019)\citenamefont {Cao},
  \citenamefont {Romero}, \citenamefont {Olson}, \citenamefont {Degroote},
  \citenamefont {Johnson}, \citenamefont {Kieferov{\'a}}, \citenamefont
  {Kivlichan}, \citenamefont {Menke}, \citenamefont {Peropadre}, \citenamefont
  {Sawaya}, \citenamefont {Sim}, \citenamefont {Veis},\ and\ \citenamefont
  {{Aspuru-Guzik}}}]{cao2019}%
  \BibitemOpen
  \bibfield  {author} {\bibinfo {author} {\bibfnamefont {Y.}~\bibnamefont
  {Cao}}, \bibinfo {author} {\bibfnamefont {J.}~\bibnamefont {Romero}},
  \bibinfo {author} {\bibfnamefont {J.~P.}\ \bibnamefont {Olson}}, \bibinfo
  {author} {\bibfnamefont {M.}~\bibnamefont {Degroote}}, \bibinfo {author}
  {\bibfnamefont {P.~D.}\ \bibnamefont {Johnson}}, \bibinfo {author}
  {\bibfnamefont {M.}~\bibnamefont {Kieferov{\'a}}}, \bibinfo {author}
  {\bibfnamefont {I.~D.}\ \bibnamefont {Kivlichan}}, \bibinfo {author}
  {\bibfnamefont {T.}~\bibnamefont {Menke}}, \bibinfo {author} {\bibfnamefont
  {B.}~\bibnamefont {Peropadre}}, \bibinfo {author} {\bibfnamefont {N.~P.~D.}\
  \bibnamefont {Sawaya}}, \bibinfo {author} {\bibfnamefont {S.}~\bibnamefont
  {Sim}}, \bibinfo {author} {\bibfnamefont {L.}~\bibnamefont {Veis}},\ and\
  \bibinfo {author} {\bibfnamefont {A.}~\bibnamefont {{Aspuru-Guzik}}},\
  }\bibfield  {title} {\bibinfo {title} {Quantum {{Chemistry}} in the {{Age}}
  of {{Quantum Computing}}},\ }\href
  {https://doi.org/10.1021/acs.chemrev.8b00803} {\bibfield  {journal} {\bibinfo
   {journal} {Chemical Reviews}\ }\textbf {\bibinfo {volume} {119}},\ \bibinfo
  {pages} {10856} (\bibinfo {year} {2019})}\BibitemShut {NoStop}%
\bibitem [{\citenamefont {Bauer}\ \emph {et~al.}(2020)\citenamefont {Bauer},
  \citenamefont {Bravyi}, \citenamefont {Motta},\ and\ \citenamefont
  {Chan}}]{bauer2020}%
  \BibitemOpen
  \bibfield  {author} {\bibinfo {author} {\bibfnamefont {B.}~\bibnamefont
  {Bauer}}, \bibinfo {author} {\bibfnamefont {S.}~\bibnamefont {Bravyi}},
  \bibinfo {author} {\bibfnamefont {M.}~\bibnamefont {Motta}},\ and\ \bibinfo
  {author} {\bibfnamefont {G.~K.-L.}\ \bibnamefont {Chan}},\ }\bibfield
  {title} {\bibinfo {title} {Quantum {{Algorithms}} for {{Quantum Chemistry}}
  and {{Quantum Materials Science}}},\ }\href
  {https://doi.org/10.1021/acs.chemrev.9b00829} {\bibfield  {journal} {\bibinfo
   {journal} {Chemical Reviews}\ }\textbf {\bibinfo {volume} {120}},\ \bibinfo
  {pages} {12685} (\bibinfo {year} {2020})}\BibitemShut {NoStop}%
\bibitem [{\citenamefont {Babbush}\ \emph {et~al.}(2018)\citenamefont
  {Babbush}, \citenamefont {Gidney}, \citenamefont {Berry}, \citenamefont
  {Wiebe}, \citenamefont {McClean}, \citenamefont {Paler}, \citenamefont
  {Fowler},\ and\ \citenamefont {Neven}}]{babbush2018}%
  \BibitemOpen
  \bibfield  {author} {\bibinfo {author} {\bibfnamefont {R.}~\bibnamefont
  {Babbush}}, \bibinfo {author} {\bibfnamefont {C.}~\bibnamefont {Gidney}},
  \bibinfo {author} {\bibfnamefont {D.~W.}\ \bibnamefont {Berry}}, \bibinfo
  {author} {\bibfnamefont {N.}~\bibnamefont {Wiebe}}, \bibinfo {author}
  {\bibfnamefont {J.}~\bibnamefont {McClean}}, \bibinfo {author} {\bibfnamefont
  {A.}~\bibnamefont {Paler}}, \bibinfo {author} {\bibfnamefont
  {A.}~\bibnamefont {Fowler}},\ and\ \bibinfo {author} {\bibfnamefont
  {H.}~\bibnamefont {Neven}},\ }\bibfield  {title} {\bibinfo {title} {Encoding
  {{Electronic Spectra}} in {{Quantum Circuits}} with {{Linear T
  Complexity}}},\ }\href {https://doi.org/10.1103/PhysRevX.8.041015} {\bibfield
   {journal} {\bibinfo  {journal} {Physical Review X}\ }\textbf {\bibinfo
  {volume} {8}},\ \bibinfo {pages} {041015} (\bibinfo {year}
  {2018})}\BibitemShut {NoStop}%
\bibitem [{\citenamefont {Giovannetti}\ \emph
  {et~al.}(2008{\natexlab{b}})\citenamefont {Giovannetti}, \citenamefont
  {Lloyd},\ and\ \citenamefont {Maccone}}]{giovannetti2008}%
  \BibitemOpen
  \bibfield  {author} {\bibinfo {author} {\bibfnamefont {V.}~\bibnamefont
  {Giovannetti}}, \bibinfo {author} {\bibfnamefont {S.}~\bibnamefont {Lloyd}},\
  and\ \bibinfo {author} {\bibfnamefont {L.}~\bibnamefont {Maccone}},\
  }\bibfield  {title} {\bibinfo {title} {Architectures for a quantum random
  access memory},\ }\href {https://doi.org/10.1103/PhysRevA.78.052310}
  {\bibfield  {journal} {\bibinfo  {journal} {Physical Review A}\ }\textbf
  {\bibinfo {volume} {78}},\ \bibinfo {pages} {052310} (\bibinfo {year}
  {2008}{\natexlab{b}})}\BibitemShut {NoStop}%
\bibitem [{\citenamefont {Park}\ \emph {et~al.}(2019)\citenamefont {Park},
  \citenamefont {Petruccione},\ and\ \citenamefont {Rhee}}]{park2019}%
  \BibitemOpen
  \bibfield  {author} {\bibinfo {author} {\bibfnamefont {D.~K.}\ \bibnamefont
  {Park}}, \bibinfo {author} {\bibfnamefont {F.}~\bibnamefont {Petruccione}},\
  and\ \bibinfo {author} {\bibfnamefont {J.-K.~K.}\ \bibnamefont {Rhee}},\
  }\bibfield  {title} {\bibinfo {title} {Circuit-{{Based Quantum Random Access
  Memory}} for {{Classical Data}}},\ }\href
  {https://doi.org/10.1038/s41598-019-40439-3} {\bibfield  {journal} {\bibinfo
  {journal} {Scientific Reports}\ }\textbf {\bibinfo {volume} {9}},\ \bibinfo
  {pages} {3949} (\bibinfo {year} {2019})}\BibitemShut {NoStop}%
\bibitem [{\citenamefont {Matteo}\ \emph {et~al.}(2020)\citenamefont {Matteo},
  \citenamefont {Gheorghiu},\ and\ \citenamefont {Mosca}}]{matteo2020}%
  \BibitemOpen
  \bibfield  {author} {\bibinfo {author} {\bibfnamefont {O.~D.}\ \bibnamefont
  {Matteo}}, \bibinfo {author} {\bibfnamefont {V.}~\bibnamefont {Gheorghiu}},\
  and\ \bibinfo {author} {\bibfnamefont {M.}~\bibnamefont {Mosca}},\ }\bibfield
   {title} {\bibinfo {title} {Fault-{{Tolerant Resource Estimation}} of
  {{Quantum Random-Access Memories}}},\ }\href
  {https://doi.org/10.1109/TQE.2020.2965803} {\bibfield  {journal} {\bibinfo
  {journal} {IEEE Transactions on Quantum Engineering}\ }\textbf {\bibinfo
  {volume} {1}},\ \bibinfo {pages} {1} (\bibinfo {year} {2020})}\BibitemShut
  {NoStop}%
\bibitem [{\citenamefont {Paler}\ \emph {et~al.}(2020)\citenamefont {Paler},
  \citenamefont {Oumarou},\ and\ \citenamefont {Basmadjian}}]{paler2020}%
  \BibitemOpen
  \bibfield  {author} {\bibinfo {author} {\bibfnamefont {A.}~\bibnamefont
  {Paler}}, \bibinfo {author} {\bibfnamefont {O.}~\bibnamefont {Oumarou}},\
  and\ \bibinfo {author} {\bibfnamefont {R.}~\bibnamefont {Basmadjian}},\
  }\bibfield  {title} {\bibinfo {title} {Parallelizing the queries in a
  bucket-brigade quantum random access memory},\ }\href
  {https://doi.org/10.1103/PhysRevA.102.032608} {\bibfield  {journal} {\bibinfo
   {journal} {Physical Review A}\ }\textbf {\bibinfo {volume} {102}},\ \bibinfo
  {pages} {032608} (\bibinfo {year} {2020})}\BibitemShut {NoStop}%
\bibitem [{\citenamefont {Niu}\ \emph {et~al.}(2022)\citenamefont {Niu},
  \citenamefont {Zlokapa}, \citenamefont {Broughton}, \citenamefont {Boixo},
  \citenamefont {Mohseni}, \citenamefont {Smelyanskyi},\ and\ \citenamefont
  {Neven}}]{niu2022}%
  \BibitemOpen
  \bibfield  {author} {\bibinfo {author} {\bibfnamefont {M.~Y.}\ \bibnamefont
  {Niu}}, \bibinfo {author} {\bibfnamefont {A.}~\bibnamefont {Zlokapa}},
  \bibinfo {author} {\bibfnamefont {M.}~\bibnamefont {Broughton}}, \bibinfo
  {author} {\bibfnamefont {S.}~\bibnamefont {Boixo}}, \bibinfo {author}
  {\bibfnamefont {M.}~\bibnamefont {Mohseni}}, \bibinfo {author} {\bibfnamefont
  {V.}~\bibnamefont {Smelyanskyi}},\ and\ \bibinfo {author} {\bibfnamefont
  {H.}~\bibnamefont {Neven}},\ }\bibfield  {title} {\bibinfo {title}
  {Entangling {{Quantum Generative Adversarial Networks}}},\ }\href
  {https://doi.org/10.1103/PhysRevLett.128.220505} {\bibfield  {journal}
  {\bibinfo  {journal} {Physical Review Letters}\ }\textbf {\bibinfo {volume}
  {128}},\ \bibinfo {pages} {220505} (\bibinfo {year} {2022})}\BibitemShut
  {NoStop}%
\bibitem [{\citenamefont {Jaques}\ and\ \citenamefont
  {Rattew}(2023)}]{jaques2023}%
  \BibitemOpen
  \bibfield  {author} {\bibinfo {author} {\bibfnamefont {S.}~\bibnamefont
  {Jaques}}\ and\ \bibinfo {author} {\bibfnamefont {A.~G.}\ \bibnamefont
  {Rattew}},\ }\href {https://doi.org/10.48550/arXiv.2305.10310} {\bibinfo
  {title} {{{QRAM}}: {{A Survey}} and {{Critique}}}} (\bibinfo {year} {2023}),\
  \Eprint {https://arxiv.org/abs/2305.10310} {arXiv:2305.10310 [quant-ph]}
  \BibitemShut {NoStop}%
\bibitem [{\citenamefont {Phalak}\ \emph {et~al.}(2023)\citenamefont {Phalak},
  \citenamefont {Chatterjee},\ and\ \citenamefont {Ghosh}}]{phalak2023}%
  \BibitemOpen
  \bibfield  {author} {\bibinfo {author} {\bibfnamefont {K.}~\bibnamefont
  {Phalak}}, \bibinfo {author} {\bibfnamefont {A.}~\bibnamefont {Chatterjee}},\
  and\ \bibinfo {author} {\bibfnamefont {S.}~\bibnamefont {Ghosh}},\ }\bibfield
   {title} {\bibinfo {title} {Quantum {{Random Access Memory}} for
  {{Dummies}}},\ }\href {https://doi.org/10.3390/s23177462} {\bibfield
  {journal} {\bibinfo  {journal} {Sensors}\ }\textbf {\bibinfo {volume} {23}},\
  \bibinfo {pages} {7462} (\bibinfo {year} {2023})}\BibitemShut {NoStop}%
\bibitem [{\citenamefont {Xu}\ \emph {et~al.}(2023)\citenamefont {Xu},
  \citenamefont {Hann}, \citenamefont {Foxman}, \citenamefont {Girvin},\ and\
  \citenamefont {Ding}}]{xu2023}%
  \BibitemOpen
  \bibfield  {author} {\bibinfo {author} {\bibfnamefont {S.}~\bibnamefont
  {Xu}}, \bibinfo {author} {\bibfnamefont {C.~T.}\ \bibnamefont {Hann}},
  \bibinfo {author} {\bibfnamefont {B.}~\bibnamefont {Foxman}}, \bibinfo
  {author} {\bibfnamefont {S.~M.}\ \bibnamefont {Girvin}},\ and\ \bibinfo
  {author} {\bibfnamefont {Y.}~\bibnamefont {Ding}},\ }\bibfield  {title}
  {\bibinfo {title} {Systems {{Architecture}} for {{Quantum Random Access
  Memory}}},\ }in\ \href {https://doi.org/10.1145/3613424.3614270} {\emph
  {\bibinfo {booktitle} {Proceedings of the 56th {{Annual IEEE}}/{{ACM
  International Symposium}} on {{Microarchitecture}}}}},\ \bibinfo {series and
  number} {{{MICRO}} '23}\ (\bibinfo  {publisher} {Association for Computing
  Machinery},\ \bibinfo {address} {New York, NY, USA},\ \bibinfo {year}
  {2023})\ pp.\ \bibinfo {pages} {526--538}\BibitemShut {NoStop}%
\bibitem [{\citenamefont {Hann}(2021)}]{hann2021a}%
  \BibitemOpen
  \bibfield  {author} {\bibinfo {author} {\bibfnamefont {C.}~\bibnamefont
  {Hann}},\ }\bibfield  {title} {\bibinfo {title} {Practicality of {{Quantum
  Random Access Memory}}},\ }\href@noop {} {\bibfield  {journal} {\bibinfo
  {journal} {Yale Graduate School of Arts and Sciences Dissertations}\ }
  (\bibinfo {year} {2021})}\BibitemShut {NoStop}%
\bibitem [{\citenamefont {Hann}\ \emph {et~al.}(2021)\citenamefont {Hann},
  \citenamefont {Lee}, \citenamefont {Girvin},\ and\ \citenamefont
  {Jiang}}]{hann2021}%
  \BibitemOpen
  \bibfield  {author} {\bibinfo {author} {\bibfnamefont {C.~T.}\ \bibnamefont
  {Hann}}, \bibinfo {author} {\bibfnamefont {G.}~\bibnamefont {Lee}}, \bibinfo
  {author} {\bibfnamefont {S.}~\bibnamefont {Girvin}},\ and\ \bibinfo {author}
  {\bibfnamefont {L.}~\bibnamefont {Jiang}},\ }\bibfield  {title} {\bibinfo
  {title} {Resilience of {{Quantum Random Access Memory}} to {{Generic
  Noise}}},\ }\href {https://doi.org/10.1103/PRXQuantum.2.020311} {\bibfield
  {journal} {\bibinfo  {journal} {PRX Quantum}\ }\textbf {\bibinfo {volume}
  {2}},\ \bibinfo {pages} {020311} (\bibinfo {year} {2021})}\BibitemShut
  {NoStop}%
\bibitem [{\citenamefont {Hong}\ \emph {et~al.}(2012)\citenamefont {Hong},
  \citenamefont {Xiang}, \citenamefont {Zhu}, \citenamefont {Jiang},\ and\
  \citenamefont {Wu}}]{hong2012}%
  \BibitemOpen
  \bibfield  {author} {\bibinfo {author} {\bibfnamefont {F.-Y.}\ \bibnamefont
  {Hong}}, \bibinfo {author} {\bibfnamefont {Y.}~\bibnamefont {Xiang}},
  \bibinfo {author} {\bibfnamefont {Z.-Y.}\ \bibnamefont {Zhu}}, \bibinfo
  {author} {\bibfnamefont {L.-z.}\ \bibnamefont {Jiang}},\ and\ \bibinfo
  {author} {\bibfnamefont {L.-n.}\ \bibnamefont {Wu}},\ }\bibfield  {title}
  {\bibinfo {title} {Robust quantum random access memory},\ }\href
  {https://doi.org/10.1103/PhysRevA.86.010306} {\bibfield  {journal} {\bibinfo
  {journal} {Physical Review A}\ }\textbf {\bibinfo {volume} {86}},\ \bibinfo
  {pages} {010306} (\bibinfo {year} {2012})}\BibitemShut {NoStop}%
\bibitem [{\citenamefont {Moiseev}\ and\ \citenamefont
  {Moiseev}(2016)}]{moiseev2016}%
  \BibitemOpen
  \bibfield  {author} {\bibinfo {author} {\bibfnamefont {E.~S.}\ \bibnamefont
  {Moiseev}}\ and\ \bibinfo {author} {\bibfnamefont {S.~A.}\ \bibnamefont
  {Moiseev}},\ }\bibfield  {title} {\bibinfo {title} {Time-bin quantum
  {{RAM}}},\ }\href {https://doi.org/10.1080/09500340.2016.1182222} {\bibfield
  {journal} {\bibinfo  {journal} {Journal of Modern Optics}\ }\textbf {\bibinfo
  {volume} {63}},\ \bibinfo {pages} {2081} (\bibinfo {year}
  {2016})}\BibitemShut {NoStop}%
\bibitem [{\citenamefont {Weiss}\ \emph {et~al.}(2024)\citenamefont {Weiss},
  \citenamefont {Puri},\ and\ \citenamefont {Girvin}}]{weiss2024}%
  \BibitemOpen
  \bibfield  {author} {\bibinfo {author} {\bibfnamefont {D.}~\bibnamefont
  {Weiss}}, \bibinfo {author} {\bibfnamefont {S.}~\bibnamefont {Puri}},\ and\
  \bibinfo {author} {\bibfnamefont {S.}~\bibnamefont {Girvin}},\ }\bibfield
  {title} {\bibinfo {title} {Quantum {{Random Access Memory Architectures Using
  3D Superconducting Cavities}}},\ }\href
  {https://doi.org/10.1103/PRXQuantum.5.020312} {\bibfield  {journal} {\bibinfo
   {journal} {PRX Quantum}\ }\textbf {\bibinfo {volume} {5}},\ \bibinfo {pages}
  {020312} (\bibinfo {year} {2024})}\BibitemShut {NoStop}%
\bibitem [{\citenamefont {Chen}\ \emph {et~al.}(2021)\citenamefont {Chen},
  \citenamefont {Dai}, \citenamefont {{Errando-Herranz}}, \citenamefont
  {Lloyd},\ and\ \citenamefont {Englund}}]{chen2021}%
  \BibitemOpen
  \bibfield  {author} {\bibinfo {author} {\bibfnamefont {K.~C.}\ \bibnamefont
  {Chen}}, \bibinfo {author} {\bibfnamefont {W.}~\bibnamefont {Dai}}, \bibinfo
  {author} {\bibfnamefont {C.}~\bibnamefont {{Errando-Herranz}}}, \bibinfo
  {author} {\bibfnamefont {S.}~\bibnamefont {Lloyd}},\ and\ \bibinfo {author}
  {\bibfnamefont {D.}~\bibnamefont {Englund}},\ }\bibfield  {title} {\bibinfo
  {title} {Scalable and {{High-Fidelity Quantum Random Access Memory}} in
  {{Spin-Photon Networks}}},\ }\href
  {https://doi.org/10.1103/PRXQuantum.2.030319} {\bibfield  {journal} {\bibinfo
   {journal} {PRX Quantum}\ }\textbf {\bibinfo {volume} {2}},\ \bibinfo {pages}
  {030319} (\bibinfo {year} {2021})}\BibitemShut {NoStop}%
\bibitem [{\citenamefont {Hann}\ \emph {et~al.}(2019)\citenamefont {Hann},
  \citenamefont {Zou}, \citenamefont {Zhang}, \citenamefont {Chu},
  \citenamefont {Schoelkopf}, \citenamefont {Girvin},\ and\ \citenamefont
  {Jiang}}]{hann2019}%
  \BibitemOpen
  \bibfield  {author} {\bibinfo {author} {\bibfnamefont {C.~T.}\ \bibnamefont
  {Hann}}, \bibinfo {author} {\bibfnamefont {C.-L.}\ \bibnamefont {Zou}},
  \bibinfo {author} {\bibfnamefont {Y.}~\bibnamefont {Zhang}}, \bibinfo
  {author} {\bibfnamefont {Y.}~\bibnamefont {Chu}}, \bibinfo {author}
  {\bibfnamefont {R.~J.}\ \bibnamefont {Schoelkopf}}, \bibinfo {author}
  {\bibfnamefont {S.~M.}\ \bibnamefont {Girvin}},\ and\ \bibinfo {author}
  {\bibfnamefont {L.}~\bibnamefont {Jiang}},\ }\bibfield  {title} {\bibinfo
  {title} {Hardware-{{Efficient Quantum Random Access Memory}} with {{Hybrid
  Quantum Acoustic Systems}}},\ }\href
  {https://doi.org/10.1103/PhysRevLett.123.250501} {\bibfield  {journal}
  {\bibinfo  {journal} {Physical Review Letters}\ }\textbf {\bibinfo {volume}
  {123}},\ \bibinfo {pages} {250501} (\bibinfo {year} {2019})}\BibitemShut
  {NoStop}%
\bibitem [{\citenamefont {Naik}\ \emph {et~al.}(2017)\citenamefont {Naik},
  \citenamefont {Leung}, \citenamefont {Chakram}, \citenamefont {Groszkowski},
  \citenamefont {Lu}, \citenamefont {Earnest}, \citenamefont {McKay},
  \citenamefont {Koch},\ and\ \citenamefont {Schuster}}]{naik2017}%
  \BibitemOpen
  \bibfield  {author} {\bibinfo {author} {\bibfnamefont {R.~K.}\ \bibnamefont
  {Naik}}, \bibinfo {author} {\bibfnamefont {N.}~\bibnamefont {Leung}},
  \bibinfo {author} {\bibfnamefont {S.}~\bibnamefont {Chakram}}, \bibinfo
  {author} {\bibfnamefont {P.}~\bibnamefont {Groszkowski}}, \bibinfo {author}
  {\bibfnamefont {Y.}~\bibnamefont {Lu}}, \bibinfo {author} {\bibfnamefont
  {N.}~\bibnamefont {Earnest}}, \bibinfo {author} {\bibfnamefont {D.~C.}\
  \bibnamefont {McKay}}, \bibinfo {author} {\bibfnamefont {J.}~\bibnamefont
  {Koch}},\ and\ \bibinfo {author} {\bibfnamefont {D.~I.}\ \bibnamefont
  {Schuster}},\ }\bibfield  {title} {\bibinfo {title} {Random access quantum
  information processors using multimode circuit quantum electrodynamics},\
  }\href {https://doi.org/10.1038/s41467-017-02046-6} {\bibfield  {journal}
  {\bibinfo  {journal} {Nature Communications}\ }\textbf {\bibinfo {volume}
  {8}},\ \bibinfo {pages} {1904} (\bibinfo {year} {2017})}\BibitemShut
  {NoStop}%
\bibitem [{\citenamefont {Gao}\ \emph {et~al.}(2019)\citenamefont {Gao},
  \citenamefont {Lester}, \citenamefont {Chou}, \citenamefont {Frunzio},
  \citenamefont {Devoret}, \citenamefont {Jiang}, \citenamefont {Girvin},\ and\
  \citenamefont {Schoelkopf}}]{gao2019}%
  \BibitemOpen
  \bibfield  {author} {\bibinfo {author} {\bibfnamefont {Y.~Y.}\ \bibnamefont
  {Gao}}, \bibinfo {author} {\bibfnamefont {B.~J.}\ \bibnamefont {Lester}},
  \bibinfo {author} {\bibfnamefont {K.~S.}\ \bibnamefont {Chou}}, \bibinfo
  {author} {\bibfnamefont {L.}~\bibnamefont {Frunzio}}, \bibinfo {author}
  {\bibfnamefont {M.~H.}\ \bibnamefont {Devoret}}, \bibinfo {author}
  {\bibfnamefont {L.}~\bibnamefont {Jiang}}, \bibinfo {author} {\bibfnamefont
  {S.~M.}\ \bibnamefont {Girvin}},\ and\ \bibinfo {author} {\bibfnamefont
  {R.~J.}\ \bibnamefont {Schoelkopf}},\ }\bibfield  {title} {\bibinfo {title}
  {Entanglement of bosonic modes through an engineered exchange interaction},\
  }\href {https://doi.org/10.1038/s41586-019-0970-4} {\bibfield  {journal}
  {\bibinfo  {journal} {Nature}\ }\textbf {\bibinfo {volume} {566}},\ \bibinfo
  {pages} {509} (\bibinfo {year} {2019})}\BibitemShut {NoStop}%
\bibitem [{\citenamefont {Wang}\ \emph {et~al.}(2021)\citenamefont {Wang},
  \citenamefont {Wu}, \citenamefont {Bao}, \citenamefont {Li}, \citenamefont
  {Ma}, \citenamefont {Wang}, \citenamefont {Song}, \citenamefont {Zhang},\
  and\ \citenamefont {Duan}}]{wang2021}%
  \BibitemOpen
  \bibfield  {author} {\bibinfo {author} {\bibfnamefont {Z.}~\bibnamefont
  {Wang}}, \bibinfo {author} {\bibfnamefont {Y.}~\bibnamefont {Wu}}, \bibinfo
  {author} {\bibfnamefont {Z.}~\bibnamefont {Bao}}, \bibinfo {author}
  {\bibfnamefont {Y.}~\bibnamefont {Li}}, \bibinfo {author} {\bibfnamefont
  {C.}~\bibnamefont {Ma}}, \bibinfo {author} {\bibfnamefont {H.}~\bibnamefont
  {Wang}}, \bibinfo {author} {\bibfnamefont {Y.}~\bibnamefont {Song}}, \bibinfo
  {author} {\bibfnamefont {H.}~\bibnamefont {Zhang}},\ and\ \bibinfo {author}
  {\bibfnamefont {L.}~\bibnamefont {Duan}},\ }\bibfield  {title} {\bibinfo
  {title} {Experimental {{Realization}} of a {{Deterministic Quantum Router}}
  with {{Superconducting Quantum Circuits}}},\ }\href
  {https://doi.org/10.1103/PhysRevApplied.15.014049} {\bibfield  {journal}
  {\bibinfo  {journal} {Physical Review Applied}\ }\textbf {\bibinfo {volume}
  {15}},\ \bibinfo {pages} {014049} (\bibinfo {year} {2021})}\BibitemShut
  {NoStop}%
\bibitem [{\citenamefont {Qiao}\ \emph {et~al.}(2023)\citenamefont {Qiao},
  \citenamefont {Dumur}, \citenamefont {Andersson}, \citenamefont {Yan},
  \citenamefont {Chou}, \citenamefont {Grebel}, \citenamefont {Conner},
  \citenamefont {Joshi}, \citenamefont {Miller}, \citenamefont {Povey},
  \citenamefont {Wu},\ and\ \citenamefont {Cleland}}]{qiao2023}%
  \BibitemOpen
  \bibfield  {author} {\bibinfo {author} {\bibfnamefont {H.}~\bibnamefont
  {Qiao}}, \bibinfo {author} {\bibfnamefont {{\'E}.}~\bibnamefont {Dumur}},
  \bibinfo {author} {\bibfnamefont {G.}~\bibnamefont {Andersson}}, \bibinfo
  {author} {\bibfnamefont {H.}~\bibnamefont {Yan}}, \bibinfo {author}
  {\bibfnamefont {M.-H.}\ \bibnamefont {Chou}}, \bibinfo {author}
  {\bibfnamefont {J.}~\bibnamefont {Grebel}}, \bibinfo {author} {\bibfnamefont
  {C.~R.}\ \bibnamefont {Conner}}, \bibinfo {author} {\bibfnamefont {Y.~J.}\
  \bibnamefont {Joshi}}, \bibinfo {author} {\bibfnamefont {J.~M.}\ \bibnamefont
  {Miller}}, \bibinfo {author} {\bibfnamefont {R.~G.}\ \bibnamefont {Povey}},
  \bibinfo {author} {\bibfnamefont {X.}~\bibnamefont {Wu}},\ and\ \bibinfo
  {author} {\bibfnamefont {A.~N.}\ \bibnamefont {Cleland}},\ }\bibfield
  {title} {\bibinfo {title} {Splitting phonons: {{Building}} a platform for
  linear mechanical quantum computing},\ }\href
  {https://doi.org/10.1126/science.adg8715} {\bibfield  {journal} {\bibinfo
  {journal} {Science}\ }\textbf {\bibinfo {volume} {380}},\ \bibinfo {pages}
  {1030} (\bibinfo {year} {2023})}\BibitemShut {NoStop}%
\bibitem [{\citenamefont {Chen}\ \emph {et~al.}(2014)\citenamefont {Chen},
  \citenamefont {Neill}, \citenamefont {Roushan}, \citenamefont {Leung},
  \citenamefont {Fang}, \citenamefont {Barends}, \citenamefont {Kelly},
  \citenamefont {Campbell}, \citenamefont {Chen}, \citenamefont {Chiaro},
  \citenamefont {Dunsworth}, \citenamefont {Jeffrey}, \citenamefont {Megrant},
  \citenamefont {Mutus}, \citenamefont {O'Malley}, \citenamefont {Quintana},
  \citenamefont {Sank}, \citenamefont {Vainsencher}, \citenamefont {Wenner},
  \citenamefont {White}, \citenamefont {Geller}, \citenamefont {Cleland},\ and\
  \citenamefont {Martinis}}]{chen2014}%
  \BibitemOpen
  \bibfield  {author} {\bibinfo {author} {\bibfnamefont {Y.}~\bibnamefont
  {Chen}}, \bibinfo {author} {\bibfnamefont {C.}~\bibnamefont {Neill}},
  \bibinfo {author} {\bibfnamefont {P.}~\bibnamefont {Roushan}}, \bibinfo
  {author} {\bibfnamefont {N.}~\bibnamefont {Leung}}, \bibinfo {author}
  {\bibfnamefont {M.}~\bibnamefont {Fang}}, \bibinfo {author} {\bibfnamefont
  {R.}~\bibnamefont {Barends}}, \bibinfo {author} {\bibfnamefont
  {J.}~\bibnamefont {Kelly}}, \bibinfo {author} {\bibfnamefont
  {B.}~\bibnamefont {Campbell}}, \bibinfo {author} {\bibfnamefont
  {Z.}~\bibnamefont {Chen}}, \bibinfo {author} {\bibfnamefont {B.}~\bibnamefont
  {Chiaro}}, \bibinfo {author} {\bibfnamefont {A.}~\bibnamefont {Dunsworth}},
  \bibinfo {author} {\bibfnamefont {E.}~\bibnamefont {Jeffrey}}, \bibinfo
  {author} {\bibfnamefont {A.}~\bibnamefont {Megrant}}, \bibinfo {author}
  {\bibfnamefont {J.~Y.}\ \bibnamefont {Mutus}}, \bibinfo {author}
  {\bibfnamefont {P.~J.~J.}\ \bibnamefont {O'Malley}}, \bibinfo {author}
  {\bibfnamefont {C.~M.}\ \bibnamefont {Quintana}}, \bibinfo {author}
  {\bibfnamefont {D.}~\bibnamefont {Sank}}, \bibinfo {author} {\bibfnamefont
  {A.}~\bibnamefont {Vainsencher}}, \bibinfo {author} {\bibfnamefont
  {J.}~\bibnamefont {Wenner}}, \bibinfo {author} {\bibfnamefont {T.~C.}\
  \bibnamefont {White}}, \bibinfo {author} {\bibfnamefont {M.~R.}\ \bibnamefont
  {Geller}}, \bibinfo {author} {\bibfnamefont {A.~N.}\ \bibnamefont
  {Cleland}},\ and\ \bibinfo {author} {\bibfnamefont {J.~M.}\ \bibnamefont
  {Martinis}},\ }\bibfield  {title} {\bibinfo {title} {Qubit {{Architecture}}
  with {{High Coherence}} and {{Fast Tunable Coupling}}},\ }\href
  {https://doi.org/10.1103/PhysRevLett.113.220502} {\bibfield  {journal}
  {\bibinfo  {journal} {Physical Review Letters}\ }\textbf {\bibinfo {volume}
  {113}},\ \bibinfo {pages} {220502} (\bibinfo {year} {2014})}\BibitemShut
  {NoStop}%
\bibitem [{\citenamefont {Bienfait}\ \emph {et~al.}(2019)\citenamefont
  {Bienfait}, \citenamefont {Satzinger}, \citenamefont {Zhong}, \citenamefont
  {Chang}, \citenamefont {Chou}, \citenamefont {Conner}, \citenamefont {Dumur},
  \citenamefont {Grebel}, \citenamefont {Peairs}, \citenamefont {Povey},\ and\
  \citenamefont {Cleland}}]{bienfait2019}%
  \BibitemOpen
  \bibfield  {author} {\bibinfo {author} {\bibfnamefont {A.}~\bibnamefont
  {Bienfait}}, \bibinfo {author} {\bibfnamefont {K.~J.}\ \bibnamefont
  {Satzinger}}, \bibinfo {author} {\bibfnamefont {Y.~P.}\ \bibnamefont
  {Zhong}}, \bibinfo {author} {\bibfnamefont {H.-S.}\ \bibnamefont {Chang}},
  \bibinfo {author} {\bibfnamefont {M.-H.}\ \bibnamefont {Chou}}, \bibinfo
  {author} {\bibfnamefont {C.~R.}\ \bibnamefont {Conner}}, \bibinfo {author}
  {\bibfnamefont {{\'E}.}~\bibnamefont {Dumur}}, \bibinfo {author}
  {\bibfnamefont {J.}~\bibnamefont {Grebel}}, \bibinfo {author} {\bibfnamefont
  {G.~A.}\ \bibnamefont {Peairs}}, \bibinfo {author} {\bibfnamefont {R.~G.}\
  \bibnamefont {Povey}},\ and\ \bibinfo {author} {\bibfnamefont {A.~N.}\
  \bibnamefont {Cleland}},\ }\bibfield  {title} {\bibinfo {title}
  {Phonon-mediated quantum state transfer and remote qubit entanglement},\
  }\href {https://doi.org/10.1126/science.aaw8415} {\bibfield  {journal}
  {\bibinfo  {journal} {Science}\ }\textbf {\bibinfo {volume} {364}},\ \bibinfo
  {pages} {368} (\bibinfo {year} {2019})}\BibitemShut {NoStop}%
\bibitem [{\citenamefont {Qiao}\ and\ \citenamefont {{et al.}}()}]{qiao}%
  \BibitemOpen
  \bibfield  {author} {\bibinfo {author} {\bibfnamefont {H.}~\bibnamefont
  {Qiao}}\ and\ \bibinfo {author} {\bibnamefont {{et al.}}},\ }\bibfield
  {title} {\bibinfo {title} {Acoustic phonon phase gates with number-resolving
  phonon detection},\ }\href@noop {} {\bibinfo  {journal} {Manuscript in
  Preparation}\ }\BibitemShut {NoStop}%
\bibitem [{\citenamefont {Bienfait}\ \emph {et~al.}(2020)\citenamefont
  {Bienfait}, \citenamefont {Zhong}, \citenamefont {Chang}, \citenamefont
  {Chou}, \citenamefont {Conner}, \citenamefont {Dumur}, \citenamefont
  {Grebel}, \citenamefont {Peairs}, \citenamefont {Povey}, \citenamefont
  {Satzinger},\ and\ \citenamefont {Cleland}}]{bienfait2020}%
  \BibitemOpen
\bibfield  {journal} {  }\bibfield  {author} {\bibinfo {author} {\bibfnamefont
  {A.}~\bibnamefont {Bienfait}}, \bibinfo {author} {\bibfnamefont {Y.~P.}\
  \bibnamefont {Zhong}}, \bibinfo {author} {\bibfnamefont {H.-S.}\ \bibnamefont
  {Chang}}, \bibinfo {author} {\bibfnamefont {M.-H.}\ \bibnamefont {Chou}},
  \bibinfo {author} {\bibfnamefont {C.~R.}\ \bibnamefont {Conner}}, \bibinfo
  {author} {\bibfnamefont {{\'E}.}~\bibnamefont {Dumur}}, \bibinfo {author}
  {\bibfnamefont {J.}~\bibnamefont {Grebel}}, \bibinfo {author} {\bibfnamefont
  {G.~A.}\ \bibnamefont {Peairs}}, \bibinfo {author} {\bibfnamefont {R.~G.}\
  \bibnamefont {Povey}}, \bibinfo {author} {\bibfnamefont {K.~J.}\ \bibnamefont
  {Satzinger}},\ and\ \bibinfo {author} {\bibfnamefont {A.~N.}\ \bibnamefont
  {Cleland}},\ }\bibfield  {title} {\bibinfo {title} {Quantum {{Erasure Using
  Entangled Surface Acoustic Phonons}}},\ }\href
  {https://doi.org/10.1103/PhysRevX.10.021055} {\bibfield  {journal} {\bibinfo
  {journal} {Physical Review X}\ }\textbf {\bibinfo {volume} {10}},\ \bibinfo
  {pages} {021055} (\bibinfo {year} {2020})}\BibitemShut {NoStop}%
\bibitem [{\citenamefont {Satzinger}\ \emph {et~al.}(2018)\citenamefont
  {Satzinger}, \citenamefont {Zhong}, \citenamefont {Chang}, \citenamefont
  {Peairs}, \citenamefont {Bienfait}, \citenamefont {Chou}, \citenamefont
  {Cleland}, \citenamefont {Conner}, \citenamefont {Dumur}, \citenamefont
  {Grebel}, \citenamefont {Gutierrez}, \citenamefont {November}, \citenamefont
  {Povey}, \citenamefont {Whiteley}, \citenamefont {Awschalom}, \citenamefont
  {Schuster},\ and\ \citenamefont {Cleland}}]{satzinger2018}%
  \BibitemOpen
  \bibfield  {author} {\bibinfo {author} {\bibfnamefont {K.~J.}\ \bibnamefont
  {Satzinger}}, \bibinfo {author} {\bibfnamefont {Y.~P.}\ \bibnamefont
  {Zhong}}, \bibinfo {author} {\bibfnamefont {H.-S.}\ \bibnamefont {Chang}},
  \bibinfo {author} {\bibfnamefont {G.~A.}\ \bibnamefont {Peairs}}, \bibinfo
  {author} {\bibfnamefont {A.}~\bibnamefont {Bienfait}}, \bibinfo {author}
  {\bibfnamefont {M.-H.}\ \bibnamefont {Chou}}, \bibinfo {author}
  {\bibfnamefont {A.~Y.}\ \bibnamefont {Cleland}}, \bibinfo {author}
  {\bibfnamefont {C.~R.}\ \bibnamefont {Conner}}, \bibinfo {author}
  {\bibfnamefont {{\'E}.}~\bibnamefont {Dumur}}, \bibinfo {author}
  {\bibfnamefont {J.}~\bibnamefont {Grebel}}, \bibinfo {author} {\bibfnamefont
  {I.}~\bibnamefont {Gutierrez}}, \bibinfo {author} {\bibfnamefont {B.~H.}\
  \bibnamefont {November}}, \bibinfo {author} {\bibfnamefont {R.~G.}\
  \bibnamefont {Povey}}, \bibinfo {author} {\bibfnamefont {S.~J.}\ \bibnamefont
  {Whiteley}}, \bibinfo {author} {\bibfnamefont {D.~D.}\ \bibnamefont
  {Awschalom}}, \bibinfo {author} {\bibfnamefont {D.~I.}\ \bibnamefont
  {Schuster}},\ and\ \bibinfo {author} {\bibfnamefont {A.~N.}\ \bibnamefont
  {Cleland}},\ }\bibfield  {title} {\bibinfo {title} {Quantum control of
  surface acoustic-wave phonons},\ }\href
  {https://doi.org/10.1038/s41586-018-0719-5} {\bibfield  {journal} {\bibinfo
  {journal} {Nature}\ }\textbf {\bibinfo {volume} {563}},\ \bibinfo {pages}
  {661} (\bibinfo {year} {2018})}\BibitemShut {NoStop}%
\bibitem [{\citenamefont {Zivari}\ \emph {et~al.}(2021)\citenamefont {Zivari},
  \citenamefont {Stockill}, \citenamefont {Fiaschi},\ and\ \citenamefont
  {Gr{\"o}blacher}}]{zivari2021}%
  \BibitemOpen
  \bibfield  {author} {\bibinfo {author} {\bibfnamefont {A.}~\bibnamefont
  {Zivari}}, \bibinfo {author} {\bibfnamefont {R.}~\bibnamefont {Stockill}},
  \bibinfo {author} {\bibfnamefont {N.}~\bibnamefont {Fiaschi}},\ and\ \bibinfo
  {author} {\bibfnamefont {S.}~\bibnamefont {Gr{\"o}blacher}},\ }\bibfield
  {title} {\bibinfo {title} {Non-classical mechanical states guided in a
  phononic waveguide},\ }\href@noop {} {\bibfield  {journal} {\bibinfo
  {journal} {arXiv:2108.06248 [cond-mat, physics:physics, physics:quant-ph]}\ }
  (\bibinfo {year} {2021})},\ \Eprint {https://arxiv.org/abs/2108.06248}
  {arXiv:2108.06248 [cond-mat, physics:physics, physics:quant-ph]} \BibitemShut
  {NoStop}%
\bibitem [{\citenamefont {Kuzyk}\ and\ \citenamefont {Wang}(2018)}]{kuzyk2018}%
  \BibitemOpen
  \bibfield  {author} {\bibinfo {author} {\bibfnamefont {M.~C.}\ \bibnamefont
  {Kuzyk}}\ and\ \bibinfo {author} {\bibfnamefont {H.}~\bibnamefont {Wang}},\
  }\bibfield  {title} {\bibinfo {title} {Scaling {{Phononic Quantum Networks}}
  of {{Solid-State Spins}} with {{Closed Mechanical Subsystems}}},\ }\href
  {https://doi.org/10.1103/PhysRevX.8.041027} {\bibfield  {journal} {\bibinfo
  {journal} {Physical Review X}\ }\textbf {\bibinfo {volume} {8}},\ \bibinfo
  {pages} {041027} (\bibinfo {year} {2018})}\BibitemShut {NoStop}%
\bibitem [{\citenamefont {Lemonde}\ \emph {et~al.}(2018)\citenamefont
  {Lemonde}, \citenamefont {Meesala}, \citenamefont {Sipahigil}, \citenamefont
  {Schuetz}, \citenamefont {Lukin}, \citenamefont {Loncar},\ and\ \citenamefont
  {Rabl}}]{lemonde2018}%
  \BibitemOpen
  \bibfield  {author} {\bibinfo {author} {\bibfnamefont {M.-A.}\ \bibnamefont
  {Lemonde}}, \bibinfo {author} {\bibfnamefont {S.}~\bibnamefont {Meesala}},
  \bibinfo {author} {\bibfnamefont {A.}~\bibnamefont {Sipahigil}}, \bibinfo
  {author} {\bibfnamefont {M.~J.~A.}\ \bibnamefont {Schuetz}}, \bibinfo
  {author} {\bibfnamefont {M.~D.}\ \bibnamefont {Lukin}}, \bibinfo {author}
  {\bibfnamefont {M.}~\bibnamefont {Loncar}},\ and\ \bibinfo {author}
  {\bibfnamefont {P.}~\bibnamefont {Rabl}},\ }\bibfield  {title} {\bibinfo
  {title} {Phonon {{Networks}} with {{Silicon-Vacancy Centers}} in {{Diamond
  Waveguides}}},\ }\href {https://doi.org/10.1103/PhysRevLett.120.213603}
  {\bibfield  {journal} {\bibinfo  {journal} {Physical Review Letters}\
  }\textbf {\bibinfo {volume} {120}},\ \bibinfo {pages} {213603} (\bibinfo
  {year} {2018})}\BibitemShut {NoStop}%
\bibitem [{\citenamefont {Zhan}\ and\ \citenamefont {Sun}(2020)}]{zhan2020}%
  \BibitemOpen
  \bibfield  {author} {\bibinfo {author} {\bibfnamefont {Y.}~\bibnamefont
  {Zhan}}\ and\ \bibinfo {author} {\bibfnamefont {S.}~\bibnamefont {Sun}},\
  }\bibfield  {title} {\bibinfo {title} {Deterministic {{Generation}} of
  {{Loss-Tolerant Photonic Cluster States}} with a {{Single Quantum
  Emitter}}},\ }\href {https://doi.org/10.1103/PhysRevLett.125.223601}
  {\bibfield  {journal} {\bibinfo  {journal} {Physical Review Letters}\
  }\textbf {\bibinfo {volume} {125}},\ \bibinfo {pages} {223601} (\bibinfo
  {year} {2020})}\BibitemShut {NoStop}%
\bibitem [{\citenamefont {Kiilerich}\ and\ \citenamefont
  {M{\o}lmer}(2019)}]{kiilerich2019}%
  \BibitemOpen
  \bibfield  {author} {\bibinfo {author} {\bibfnamefont {A.~H.}\ \bibnamefont
  {Kiilerich}}\ and\ \bibinfo {author} {\bibfnamefont {K.}~\bibnamefont
  {M{\o}lmer}},\ }\bibfield  {title} {\bibinfo {title} {Input-{{Output Theory}}
  with {{Quantum Pulses}}},\ }\href
  {https://doi.org/10.1103/PhysRevLett.123.123604} {\bibfield  {journal}
  {\bibinfo  {journal} {Physical Review Letters}\ }\textbf {\bibinfo {volume}
  {123}},\ \bibinfo {pages} {123604} (\bibinfo {year} {2019})}\BibitemShut
  {NoStop}%
\bibitem [{\citenamefont {Bianchetti}\ \emph {et~al.}(2010)\citenamefont
  {Bianchetti}, \citenamefont {Filipp}, \citenamefont {Baur}, \citenamefont
  {Fink}, \citenamefont {Lang}, \citenamefont {Steffen}, \citenamefont
  {Boissonneault}, \citenamefont {Blais},\ and\ \citenamefont
  {Wallraff}}]{bianchetti2010}%
  \BibitemOpen
  \bibfield  {author} {\bibinfo {author} {\bibfnamefont {R.}~\bibnamefont
  {Bianchetti}}, \bibinfo {author} {\bibfnamefont {S.}~\bibnamefont {Filipp}},
  \bibinfo {author} {\bibfnamefont {M.}~\bibnamefont {Baur}}, \bibinfo {author}
  {\bibfnamefont {J.~M.}\ \bibnamefont {Fink}}, \bibinfo {author}
  {\bibfnamefont {C.}~\bibnamefont {Lang}}, \bibinfo {author} {\bibfnamefont
  {L.}~\bibnamefont {Steffen}}, \bibinfo {author} {\bibfnamefont
  {M.}~\bibnamefont {Boissonneault}}, \bibinfo {author} {\bibfnamefont
  {A.}~\bibnamefont {Blais}},\ and\ \bibinfo {author} {\bibfnamefont
  {A.}~\bibnamefont {Wallraff}},\ }\bibfield  {title} {\bibinfo {title}
  {Control and {{Tomography}} of a {{Three Level Superconducting Artificial
  Atom}}},\ }\href {https://doi.org/10.1103/PhysRevLett.105.223601} {\bibfield
  {journal} {\bibinfo  {journal} {Physical Review Letters}\ }\textbf {\bibinfo
  {volume} {105}},\ \bibinfo {pages} {223601} (\bibinfo {year}
  {2010})}\BibitemShut {NoStop}%
\end{thebibliography}%

\end{document}